\documentclass{IEEEtran}

\usepackage{mypreamble}
\newcommand{\SystemName}{\text{CASTER}}

\begin{document}

\title{{{\SystemName}: A Computer-Vision-Assisted Wireless Channel Simulator for Gesture Recognition}}

\author{Zhenyu Ren, Guoliang Li, Chenqing Ji, Shuai Wang, Chao Yu, Rui Wang}

% \author{Zhenyu Ren\IEEEauthorrefmark{1}, Guoliang Li\IEEEauthorrefmark{2} \IEEEmembership{(Graduate Student Member, IEEE)}, Chenqing Ji\IEEEauthorrefmark{1} \IEEEmembership{(Student Member, IEEE)}, Chao Yu\IEEEauthorrefmark{1}, Shuai Wang\IEEEauthorrefmark{3} \IEEEmembership{(Member, IEEE)}, Rui Wang\IEEEauthorrefmark{1} \IEEEmembership{(Member, IEEE)}
% \affil{Department of Electronic and Electrical Engineering, Southern University of Science and Technology (SUSTech), Shenzhen 518055, China}
% \affil{State Key Laboratory of Internet of Things for Smart City (SKL-IOTSC) and Department of Computer and Information Science, University of Macau, Taipa, Macau 999078, China}
% \affil{Shenzhen Institute of Advanced Technology, Chinese Academy of Sciences, Shenzhen 518055, China}
% \corresp{CORRESPONDING AUTHOR: Rui Wang (e-mail: wang.r@sustech.edu.cn)}}

\maketitle

\begin{abstract}
 In this paper, a computer-vision-assisted simulation method is proposed to address the issue of training dataset acquisition for wireless hand gesture recognition. In the existing literature, in order to classify gestures via the wireless channel estimation, massive training samples should be measured in a consistent environment, consuming significant efforts. In the proposed {\SystemName} simulator, however, the training dataset can be simulated via existing videos. Particularly, in the channel simulation, a gesture is represented by a sequence of snapshots, and the channel impulse response of each snapshot is calculated via tracing the rays scattered off a primitive-based hand model. Moreover, {\SystemName} simulator relies on the existing video clips to extract the motion data of gestures. Thus, the massive measurements of wireless channel can be eliminated. The experiments first demonstrate an $83.0\%$ average recognition accuracy of simulation-to-reality inference in recognizing $5$ categories of gestures. Moreover, this accuracy can be boosted to $96.5\%$ via the method of transfer learning. 
\end{abstract}


\begin{figure*}[t]
    \centering
    \includegraphics[width=\linewidth]{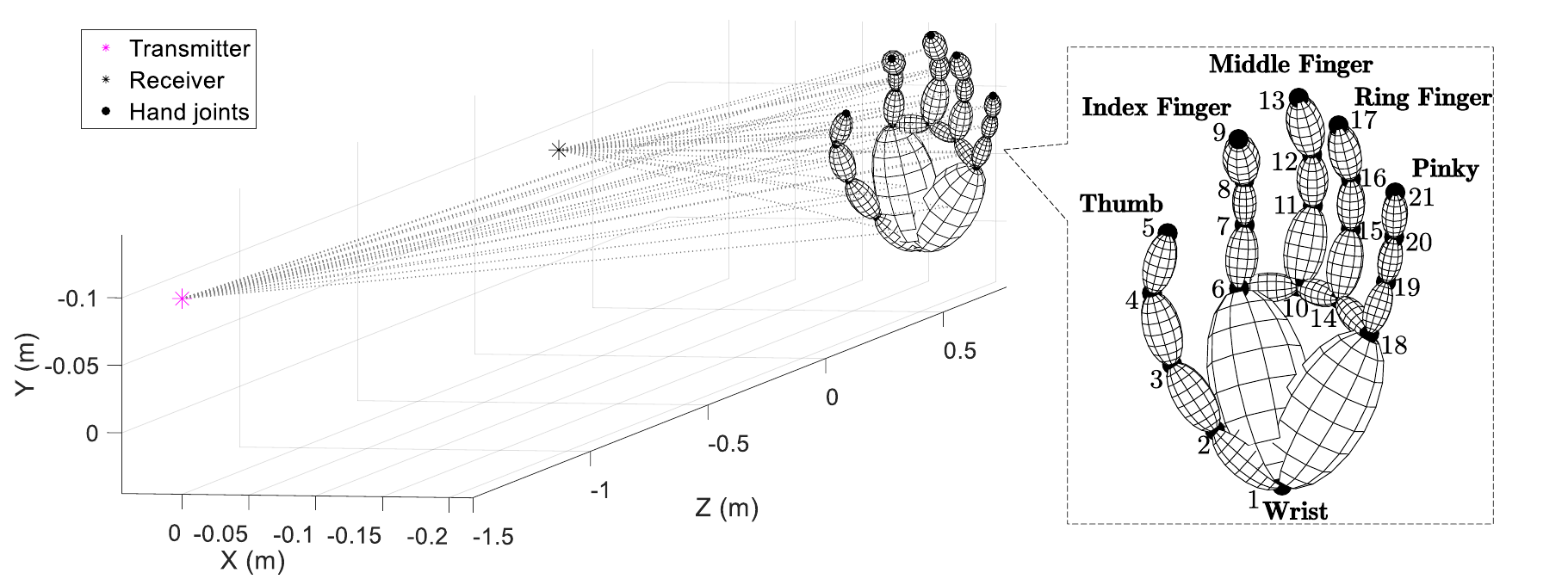}
    \caption{Illustration of primitive-based hand model and channel simulation scenario.}
    \label{fig:framework}
    % \vspace{-0.5cm}
\end{figure*}

\section{Introduction}
\label{sec:introduction}

Sensing is becoming one of the core services of the  next-generation wireless systems. There have been a significant number of works on wireless sensing, particularly the machine-learning-based human motion recognition (HMR), via channel state information (CSI)\cite{ma2019wifi,wireless_sensing_survey,zhang2021widar3, 9321398} or passive  architecture\cite{li2020passive,sun2021through,li2022passive}. In most of these works, a significant number of labeled wireless signals should be collected and processed for the training of motion recognition models, which might be infeasible in many applications. In this paper, we would like to show that it is possible to generate the above training dataset for hand gesture recognition via channel simulation, instead of real measurement.

In fact, there have been a number of works on the  extension of sophisticated communication channel models, such that the effects of sensing target on the channel impulse response are incorporated. Hence, the channel simulation based on these models might be used for motion recognition. For instance, the Data-Driven Hybrid Channel (DAHC) model of IEEE 802.11bf specification \cite{du2023overview,zhang2021channel} divided wireless channel into two parts: 
the target-unrelated components and the target-related components. The existing methods of communication channel modeling can be applied on the former; whereas the primitive-based human body model \cite{PBAH} was utilized to compute the latter. A similar channel model was also used in \cite{li2023integrated} for the optimization of communication and sensing performance. The \textit{WiGig Tools}  \cite{wigigtools}, developed by National Institute of Standards and Technology (NIST), enriched existing quasi-deterministic channel ray-tracers with supplementary target-related rays (T-Rays), such that the consistent effects of human motion could be included. Moreover, the methods for simulating radar echo signals off the human body were proposed in \cite{vishwakarma2021simhumalator,kinectDoppler}. All the above works relied on the primitive-based human body model \cite{PBAH,boulic1990global}, where the hand was modeled as a single ellipsoid. Thus, these methods cannot model fine-grained hand gestures. 

In order to facilitate the machine-learning-based HMR with the above channel models, diversified motion data are required to drive the primitive-based human body model in the channel simulation. Depth cameras and wearable sensors were used in \cite{vishwakarma2021simhumalator,kinectDoppler} to obtain sufficient body motion data for channel simulation. Nevertheless, to the best of our knowledge, there is no study on the capture of hand gestures for channel simulation. Moreover, it is unknown if conventional monocular cameras, instead of depth cameras, could obtain the motion data with adequate accuracy in the applications of wireless HMR. Note that the monocular cameras are of lower cost, and it is much more convenient to obtain hand gesture video clips of monocular cameras from online sources.  

In this paper, we would like to shed some light on the above issues by proposing a {C}omputer-vision-{A}ssisted wireless channel {S}imula{T}or for g{E}sture {R}ecognition, namely {\SystemName}. The proposed {\SystemName} simulator is composed of channel generator and video gesture catcher. In the channel generator, the target hand is modeled with $21$ primitives, and the channel impulse response is calculated by tracing the rays scattered off all the primitives. Based on the hand model, a gesture is represented by a sequence of snapshots, and the channel impulse responses for all the snapshots can be obtained respectively. In the video gesture catcher, trajectories of $21$ primitives in one gesture can be captured from videos of a conventional monocular camera. Thus, the catcher provides an efficient way to retrieve motion data for the channel generator. In order to demonstrate the high fidelity of the proposed {\SystemName} simulator, we use the simulated dataset of channel impulse responses to train a gesture recognition model and use a passive sensing system\cite{li2022passive} to measure the real channel for model testing. It is shown that an {$83.0\%$} average recognition accuracy of simulation-to-reality inference can be achieved by recognizing $5$ categories of gestures. Moreover, this accuracy can be boosted to $96.5\%$ via the method of transfer learning\cite{9134370}, where the gesture recognition model trained via a simulated dataset is further fine-tuned with a small amount of unlabeled real measurements according to the adversarial discriminative domain adaptation (ADDA) method in \cite{8099799}. The main advantages of the proposed {\SystemName} simulator are summarized below:
\begin{itemize}
    \item Conventional measurement of training dataset for wireless HMR is replaced by channel simulation and gesture video recognition, saving the significant cost of real experiments.
    \item In the proposed {\SystemName} simulator, the locations of the signal transmitter, sensing receiver, target hand, and scattering clusters can be adjusted freely to adapt to heterogeneous sensing scenarios. 
\end{itemize}
As a result, the proposed {\SystemName} simulator has the potential to customize the gesture recognition models for heterogeneous scenarios without real measurements.

The remainder of this paper is organized as follows. The simulator framework is elaborated in Section~\ref{sec:overview}. The channel generator is presented in Section~\ref{sec:channel generator}, and the video gesture catcher is presented in Section~\ref{sec:hand kinematics converter}. The performance of the {\SystemName} simulator is evaluated in Section~\ref{sec:result}. Finally, the conclusion is drawn in Section~\ref{sec:conclusion}.

In this paper, we use the following notations:
non-bold letters are used to denote scalar values,
bold lowercase letters (e.g., $\mathbf{a}$) are used to denote column vectors,
bold uppercase letters (e.g., $\mathbf{A}$) are used to denote matrices,
$|{\mathbf{a}}|$ and $\mathbf{a}^T$ denote the L2-norm and transpose of vector $\mathbf{a}$.

\section{Simulator Framework}
\label{sec:overview}

The proposed {\SystemName} simulator is developed with the primitive-based hand model. In order to extract high-fidelity channel impulse responses from existing videos, the {\SystemName} simulator is composed of the channel generator and video gesture catcher. The former generates a sequence of channel impulse response snapshots given arbitrary hand gestures and arbitrary locations of the transmitter and receiver. The latter captures the parameters of real hand motions from existing videos as the former's input. As a result, the {\SystemName} simulator is able to provide datasets for the training of the hand gesture recognition model without real channel measurement. 

As depicted in Fig.~\ref{fig:framework}, the locations of the transmitter, receiver and the target hand can be arbitrary in the channel generator. A gesture is represented as a sequence of snapshots, with an interval of $\Delta t_\text{s}$ seconds. In each snapshot, the channel is assumed to be quasi-static, and the channel impulse response is calculated via the primitive-based method \cite{PBAH}. Particularly, the hand is modeled via $21$ keypoints (joints) and  $21$ ellipsoids (primitives) connecting the keypoints. The non-line-of-sight (NLoS) channel components via the hand can be approximated by the $21$ rays respectively scattered off the centers of all primitives. Hence, the channel impulse response of one snapshot can be obtained by aggregating all the rays from the transmitter to the receiver, including the line-of-sight (LoS) ray, the NLoS ones scattered off the target hand, and the others scattered at the environment. 

As a remark notice, the $21$-keypoint hand model is widely recognized in the fields of computer vision and biomedical engineering \cite{wheatland2015state}. The renowned hand models, such as \textit{openpose}\cite{8765346}, \textit{mediaipipe}\cite{zhang2020mediapipe}, and \textit{MANO}\cite{MANO:SIGGRAPHASIA:2017}, are all based on this $21$-keypoint representation. It could provide the same degrees of freedom in describing the complex hand and finger motions as explained in \cite{wheatland2015state}: a human hand consists of $21$ joints, yielding $27$ degrees of freedom, which are the same as the $21$-keypoint hand model.

Moreover, the proposed video gesture catcher first extracts the 3-dimensional (3D) coordinates of hand keypoints from each video frame in a local hand world coordination system via machine learning technique, converts the trajectories of the keypoints from the local hand world coordinate system to a global camera coordinate system and then eliminates the fake hops and jitters of trajectories via low-pass filtering. Finally, since the interval between two video frames, denoted as $\Delta t_\text{v}$, is usually much larger than $\Delta t_\text{s}$, an interpolation is necessary to fill a sufficient number of snapshots between two video frames. 
As a remark, the video clips for the gesture catcher can be recorded in arbitrary environment as long as the desired hand gestures can be identified by the gesture catcher. Hence, they could be obtained from massive online sources.

\section{Channel Generator}
\label{sec:channel generator}

Without loss of generality, the generation of channel impulse response for the $t$-th snapshot ($\forall t$) is elaborated in this section. As shown in Fig.~\ref{fig:framework}, the rays from the transmitter to the receiver can be categorized into two parts: target-unrelated components and target-related components. The former refers to the LoS ray and the NLoS rays scattered at the static environment, and the latter refers to the NLoS rays scattered off the target hand. Particularly, let $h(\tau,t)$ and $u(\tau,t)$ be the overall channel impulse response and target-related channel impulse response of the $t$-th snapshot, $v(\tau)$ be time-invariant target-unrelated channel impulse response. Following the channel model in \cite{zhang2021channel}, we have
\begin{align}
    h(\tau,t) = u(\tau,t) + v(\tau) \label{eq:channel},
\end{align}
where the generation of $u(\tau,t)$ and $v(\tau)$ is elaborated in the following parts respectively.

\subsection{Target-Related Channel Components}
% \begin{figure}[t]
%     \centering
%     \includegraphics[width = 0.3\linewidth]{img/hand_model-eps-converted-to.pdf}
%     \caption{21-keypoint hand model with ellipsoid RCS connections.}
%     \label{fig:hand_model}
% \end{figure}
Let $\mathbf{p}_\text{t}$ and $\mathbf{p}_\text{r}$ be the coordinates of the transmitter and the receiver respectively, $\mathbf{p}_{i}(t)$ and $\mathbf{p}_{j}(t)$ be the coordinates of the two joints associated with the $n$-th primitive in the $t$-th snapshot ($\forall n,t$). Hence, the center of the $n$-th primitive is $\mathbf{p}_n^c(t)=[\mathbf{p}_{i}(t)+\mathbf{p}_{j}(t)]/2$. As previously mentioned, each primitive is modeled as an ellipsoid, the length of the axis connecting the two joints is denoted as $2 l_n(t)$, where
\begin{equation}
    l_n (t) = |\mathbf{p}_{i}(t)-\mathbf{p}_{j}(t)|/2.
\end{equation}
Moreover, the lengths of the other two axes are identical, denoted as $2r_n(t)$. Usually, $r_n(t) < l_n(t)$, and we choose $r_n(t)=l_n(t)/2$. Hence, we shall refer to the axis connecting the two joints as the long axis of the ellipsoid. As a remark note that the primitive size ($r_n$ and $l_n$) varies slightly over time due to the non-rigid nature of human motion. 

% Notice that an ellipsoid with the x-axis, y-axis, z-axis as the axes and $r_n(t)$, $r_n(t)$, $l_n(t)$ as the lengths of semi-axes can be represented as 
% \begin{align}\label{eqn:ellipsoid}
%     (x^\prime)^2/r_n^2(t)+(y^\prime)^2/r_n^2(t)+(z^\prime)^2/l_n^2(t) = 1.
% \end{align}
% The surface of the $n$-th primitive can be obtained by rotating and translating the above ellipsoid. Let 
% \begin{align}
%     \mathbf{R}_n(t)=[\boldsymbol{u}_n(t)\ \boldsymbol{w}_n(t)\ \boldsymbol{v}_n(t)] = [...]
% \end{align}
% be the rotation matrix, the surface of the $n$-th primitive is given by
% \begin{align}
%     [x\ y\ z]^T = \mathbf{R}_n(t) [x^\prime\ y^\prime\ z^\prime]^T + \mathbf{p}^\text{c}_n(t),
% \end{align}
% where $(x^\prime,y^\prime,z^\prime)$ satisfies (\ref{eqn:ellipsoid}).

\begin{figure}[t]
    \centering
    \includegraphics[width = \linewidth]{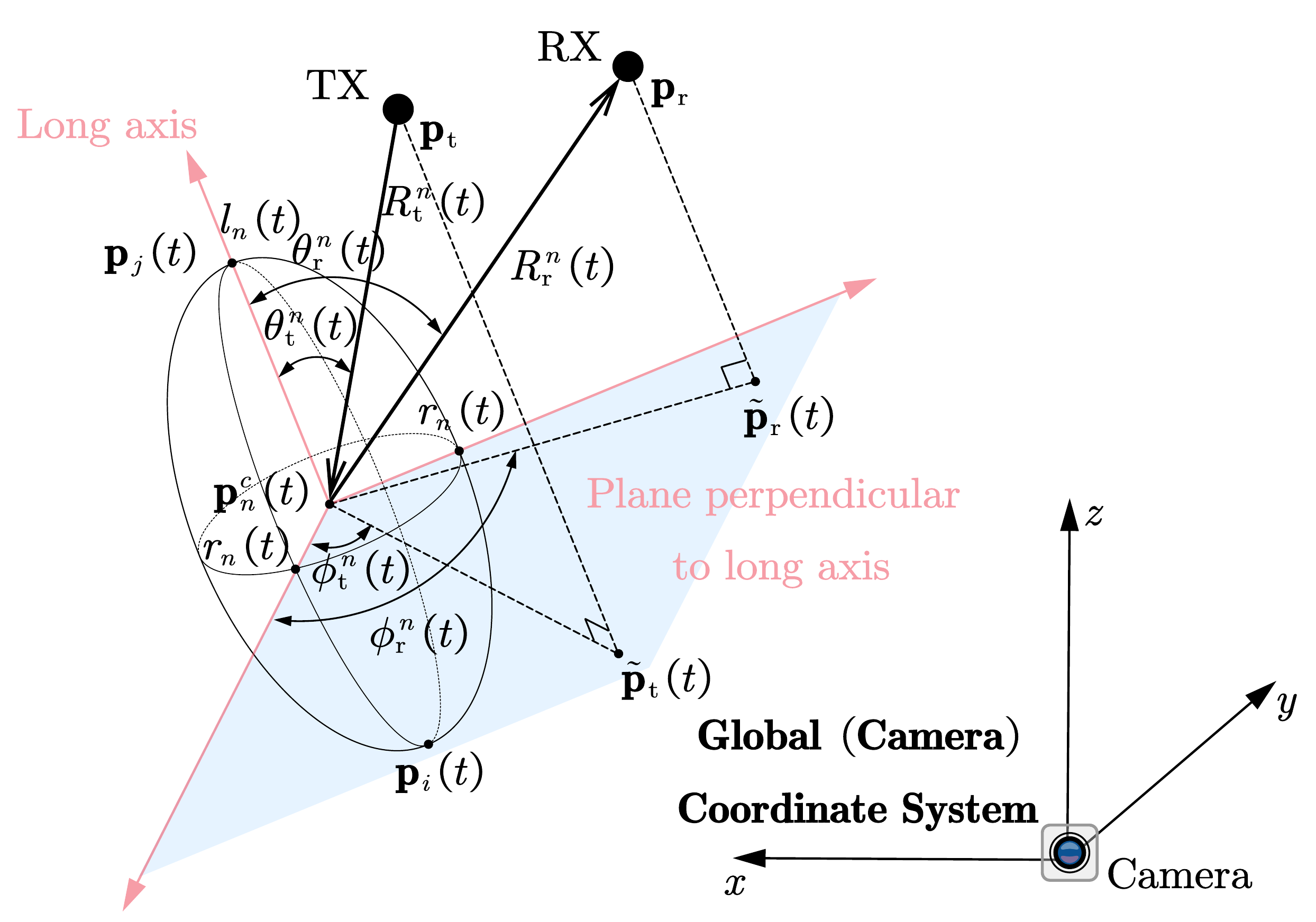}
    \caption{Bistatic RCS estimation for the $n$-th primitive.}
    \label{fig:ellipsoid_rcs}
    % \vspace{-0.3cm}
\end{figure}

Let $R_\text{t}^n(t)=|\mathbf{p}_\text{t}-\mathbf{p}^\text{c}_n(t)|$ be the distance between the transmitter and the $n$-th primitive center, $R_\text{r}^n(t)=|\mathbf{p}_\text{r}-\mathbf{p}^\text{c}_n(t)|$ be the distance between the receiver and the $n$-th primitive center, $G_\text{t}^n(t)$ and $G_\text{r}^n(t)$ be the transmit and receive antenna gains at the directions of incident ray $\mathbf{p}_\text{t}-\mathbf{p}^\text{c}_n(t)$ and scattered ray $\mathbf{p}^\text{c}_n(t)-\mathbf{p}_\text{r}$, $\sigma_n(t)$ be the bistatic radar cross section (RCS) of the $n$-th primitive, $c$ be the speed of light, $f_\text{c}$ and $\lambda $ be the carrier frequency and wavelength respectively. The response of the path scattered off the $n$-th primitive can be expressed as
\begin{align}
    u_n(\tau,t) = \lambda\sqrt{\frac{\sigma_{n}(t) G^n_\text{t}(t)G^n_\text{r}(t)}{(4\pi)^3(R_{\text{t}}^{n}(t)R_{\text{r}}^{n}(t))^2}}e^{-\mathrm{j}\phi_n(t)}\delta(\tau-\tau_n(t)),
\end{align}
where $\delta(a)$ is the impulse function, whose value is $1$ when $a=0$ and $0$ otherwise, while $\tau_{n}(t)=\left[R_\text{t}^{n}(t)+R_\text{r}^{n}(t)\right]/c$ and $\phi_{n}(t) = 2\pi f_\text{c}\tau_{n}(t)$ measure the delay and phase shift. 

Moreover, the calculation of the bistatic RCS $\sigma_{n}(t)$ follows the method in \cite{knott2004radar,trott2007stationary}. As depicted in Fig.~\ref{fig:ellipsoid_rcs}, let  $\theta^n_\text{t}(t)$ and $\theta^n _\text{r}(t)$ represent the incident and scattered elevation angles respectively,  $\phi^n_\text{t}(t)$ and $\phi^n_\text{r}(t)$ represent the incident and scattered azimuth angles respectively, $\boldsymbol{v}_n(t) = [\mathbf{p}_{i}(t)-\mathbf{p}_{j}(t)]/(2 l_n(t))$ represent the normalized vector along the long axis, we have
\begin{align}
    \theta_\text{t}^n(t)= \arccos\left( (\mathbf{p}^\text{c}_n(t)-\mathbf{p}_\text{t})^T \boldsymbol{v}_n(t)/R^n_\text{t}(t) \right),
\end{align}
\begin{align}
    \theta_\text{r}^n(t)= \arccos\left( (\mathbf{p}^\text{c}_n(t)-\mathbf{p}_\text{r})^T \boldsymbol{v}_n(t)/R^n_\text{r}(t) \right),
\end{align}
and 
\begin{small}
\begin{align}
    |\phi^n_\text{r}(t)-\phi^n_\text{t}(t)|&= \arccos\left(\frac{(\mathbf{p}^\text{c}_n(t)-\tilde{\mathbf{p}}_\text{t}(t))^T(\mathbf{p}^\text{c}_n(t)-\tilde{\mathbf{p}}_\text{r}(t))}{|\mathbf{p}^\text{c}_n(t)-\tilde{\mathbf{p}}_\text{t}(t)||\mathbf{p}^\text{c}_n(t))-\tilde{\mathbf{p}}_\text{r}(t)|}\right),
    % \nonumber
\end{align}
\end{small}

\noindent where $$\tilde{\mathbf{p}}_\text{t}(t)=\mathbf{p}_\text{t}-\boldsymbol{v}_n(t)(\mathbf{p}_\text{t}-\mathbf{p}^\text{c}_n(t))^T\boldsymbol{v}_n(t)$$ and $$\tilde{\mathbf{p}}_\text{r}(t)=\mathbf{p}_\text{r}-\boldsymbol{v}_n(t)(\mathbf{p}_\text{r}-\mathbf{p}^\text{c}_n(t))^T\boldsymbol{v}_n(t)$$ denotes the projection of the transmitter and receiver's locations on the plane containing the center of the $n$-th ellipsoid and perpendicular to its long axis in the $t$-th snapshot. As a result, the bistatic RCS $\sigma_n(t)$ of $n$-th ellipsoid in the $t$-th snapshot is given by \eqref{eq:rcs}.

\begin{figure*}[t]
    \centering
    % \begin{footnotesize}
    % \begin{equation}
    %      \sigma_n(t) = \frac{4\pi a_n^2(t)b_n^2(t)c_n^2(t)[(1+\cos \theta^n_\text{t}(t)\cos\theta^n_\text{r}(t))\cos(\phi^n_\text{r}(t)-\phi^n_\text{t}(t))+\sin\theta^n_\text{t}(t)\sin\theta^n_\text{r}(t)]^2}
    %      {[a_n^2(t)(\sin\theta^n_\text{t}(t)\cos\phi^n_\text{t}(t)+\sin\theta^n_\text{r}(t)\cos\phi^n_\text{r}(t))^2+
    %      b_n^2(t)(\sin\theta^n_\text{t}(t)\sin\phi^n_\text{t}(t)+\sin\theta^n_\text{r}(t)\sin\phi^n_\text{r}(t))^2+
    %      c_n^2(t)(\cos\theta^n_\text{t}(t)+\cos\theta^n_\text{r}(t))^2]^2}.
    % \label{eq:rcs}   
    % \end{equation}
    % \end{footnotesize}
    \begin{equation}
         \sigma_n(t) = \frac{4\pi r_n^4(t)l_n^2(t)[(1+\cos \theta^n_\text{t}(t)\cos\theta^n_\text{r}(t))\cos(\phi^n_\text{r}(t)-\phi^n_\text{t}(t))+\sin\theta^n_\text{t}(t)\sin\theta^n_\text{r}(t)]^2}
         {[r_n^2(t)(\sin^2\theta^n_\text{t}(t)+\sin^2\theta^n_\text{r}(t)+2\sin\theta^n_\text{t}(t)\sin\theta^n_\text{r}(t)\cos(\phi^n_\text{r}(t)-\phi^n_\text{t}(t)))+
         l_n^2(t)(\cos\theta^n_\text{t}(t)+\cos\theta^n_\text{r}(t))^2]^2}.
    \label{eq:rcs}   
    \end{equation}
    % \vspace{-0.3cm}
    % \begin{equation}
    % h_\text{nb}(\tau,t) = 
    % \underbrace{\sum\limits_{n=0}^{21}\left( A_n(t) \alpha_{n}(t)\exp(-\mathrm{j}\phi_{n}(t))\delta (\tau - \tilde{\tau})\right)}_{\text{target-related narrowband channel } u_\text{nb}(\tau,t)}
    % + \underbrace{\sum_{k=0}^{K-1}\left(A_k\alpha_k\exp(-\mathrm{j}\phi_k)\delta(\tau - \tilde{\tau})\right)+A_\text{LoS} \exp(-\mathrm{j}\phi_\text{LoS}) \delta(\tau-\tilde{\tau})}_{\text{target-unrelated narrowband channel } v_\text{nb}(\tau)}.
    % \label{eq:channel (narrowband)}
    % \end{equation}
    \hrule
    % \vspace{-0.6cm}
\end{figure*}

Aggregating the NLoS rays scattered off all the primitives, the target-related channel impulse response can be written as
\begin{align}
    u(\tau,t)=\sum_{n=1}^{21} u_n(\tau,t).
    \label{eq:NLos target-related}
\end{align}

\subsection{Target-Unrelated Channel Components}

{\SystemName} simulator models the environment by $K$ static scatterers. Let the RCS, transmit and receive antenna gains and the distance of the $k$-th NLoS ray be $\sigma_k$, $G_\text{t}^k$, $G_\text{r}^k$, $R_\text{t}^k$, and $R_\text{r}^k$, respectively. The NLoS components of target-unrelated channel impulse response can be written as
\begin{align}
    v_\text{NLoS}(\tau) = \sum_{k=1}^{K} \lambda\sqrt{\frac{\sigma_k G_\text{t}^kG_\text{r}^k}{(4\pi)^3(R_\text{t}^kR_\text{r}^k)^2}} e^{-\mathrm{j}\phi_k}\delta(\tau - \tau_k),
    \label{eq:NLoS target-unrelated}
\end{align}
where $\tau_k=\left(R^k_\text{t} + R^k_\text{r} \right)/c$ and $\phi_k=2\pi f_\text{c} \tau_k$. 

Moreover, let transmit and receive antenna gains at the direction of LoS path be $G_\text{t, LoS}$ and $G_\text{r, LoS}$, distance between transmitter and receiver be $R_{\text{LoS}}$, the LoS component of target-related channel is modeled via the following free space model:
\begin{align}
    v_\text{LoS}(\tau) = \frac{\lambda \sqrt{G_\text{t, LoS}G_\text{r, LoS}}}{4\pi R_{\text{LoS}}} e^{-\mathrm{j}\phi_\text{LoS}} \delta(\tau-\tau_\text{LoS}),
    \label{eq:LoS}
\end{align}
where $\tau_\text{LoS}=R_{\text{LoS}}/c$ and $\phi_\text{LoS} = 2\pi f_\text{c}\tau_\text{LoS}$. As a result, according to \cite{zhang2021channel}, the target-unrelated channel impulse response can be written as 
\begin{align}
    v(\tau) =v_\text{LoS}(\tau) + v_\text{NLoS}(\tau).
\end{align}

\section{Video Gesture Catcher}
\label{sec:hand kinematics converter}
% \begin{figure}[t]
%     \centering
%     \subfloat[]{
%         \includegraphics[width=0.45\linewidth]{img/fs300-eps-converted-to.pdf}
%     }
%     \subfloat[]{
%         \includegraphics[width=0.45\linewidth]{img/fs500-eps-converted-to.pdf}
%     }
%     \\
%     \subfloat[]{
%         \includegraphics[width=0.45\linewidth]{img/fs700-eps-converted-to.pdf}
%     }
%     \caption{The simulated spectrograms for a same motion generated by {\SystemName} at snapshot rates of (a) 300, (b) 500, and (c) 700 samples per second are presented. The maximum Doppler frequency associated with this motion is 325 Hz. It's worth noting that {\SystemName} is capable of resolving Doppler frequencies of 150 Hz, 250 Hz, and 350 Hz for (a), (b), and (c) respectively. This would lead to frequency aliasing phenomenon in (a) and (b).}
%     \label{fig:interpolation}
%     \vspace{-0.3cm}
% \end{figure}

As mentioned in the previous section, the motion of the target hand is characterized by the trajectories of the $21$ keypoints in a sequence of snapshots, denoted as $\mathbf{p}_i(t), i=1,2,...,21$. We leverage the tool of  \textit{Mediapipe} \cite{zhang2020mediapipe} to extract the keypoint trajectories from videos of monocular cameras, where two issues in the conversion are addressed in this section. The \textit{Mediapipe} could localize the positions of the keypoints in each video frame. The positions are represented in the coordinate system with the origin at the hand center, namely hand world coordinate system. However, it is difficult to calculate the Doppler frequency with such coordinate system, as the hand center is moving. Hence, we first transfer the coordinates to a unified coordinate system by solving the Perspective-n-Point (PnP) problem \cite{marchand:hal-01246370}, where the fake hops on the trajectories are smoothed. Moreover, because there are 
usually $30$ video frames per second, which is not sufficient for estimating the Doppler frequencies of gesture. For example, the typical Doppler frequencies of gestures on the $60$ GHz signals are around $800$ Hz (assuming a maximum radial velocity of $4$ meters per second), which requests $1600$ snapshots per second at least. Hence, interpolation is introduced such that the channel impulse response can be generated with a shorter interval.

\subsection{Conversion of Coordinate Systems}
% \begin{figure}[t]
%     \centering
%     \includegraphics[width = \linewidth]{img/perspective_projection-eps-converted-to.pdf}
%     \caption{Illustration of PnP coordinates transformation.}
%     \label{fig:perspective}
% \end{figure}
\begin{figure*}[t]
    \centering
    \subfloat[Pixel coordinate system]{
        \includegraphics[width=0.3\linewidth]{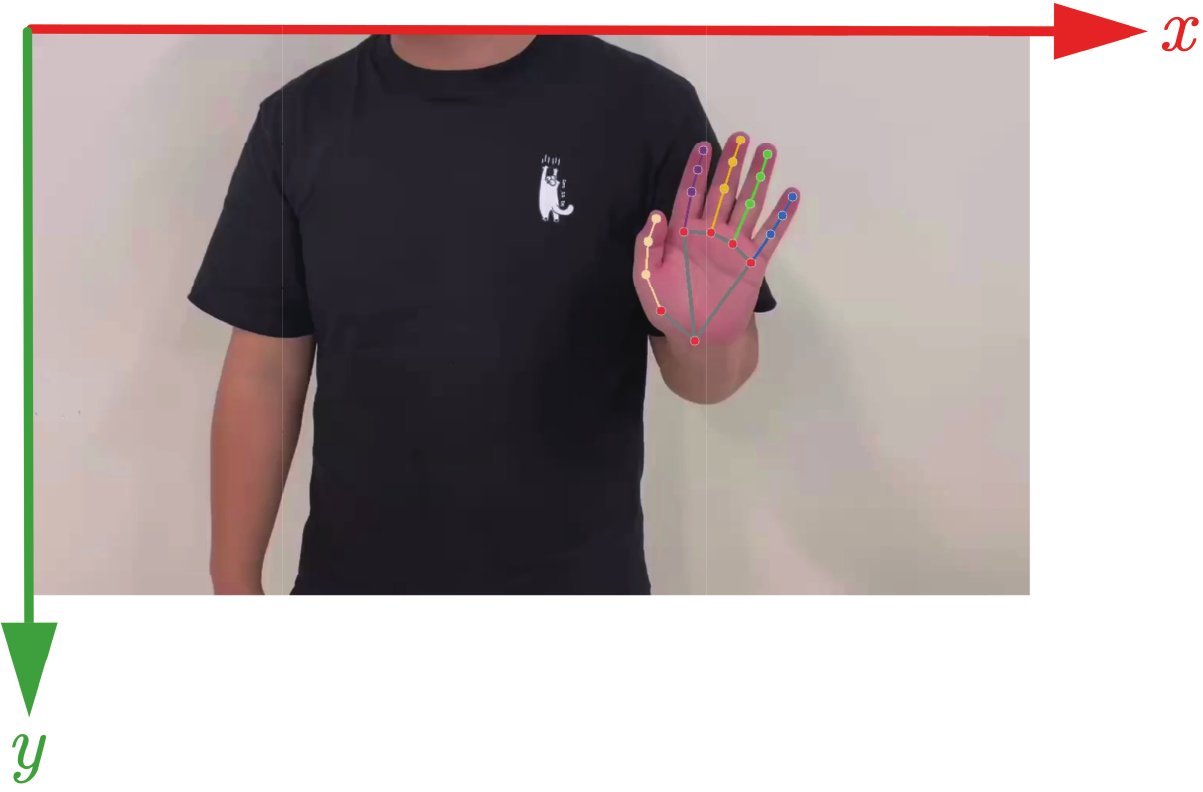}
    }
    \subfloat[Hand world coordinate system]{
        \includegraphics[width=0.3\linewidth]{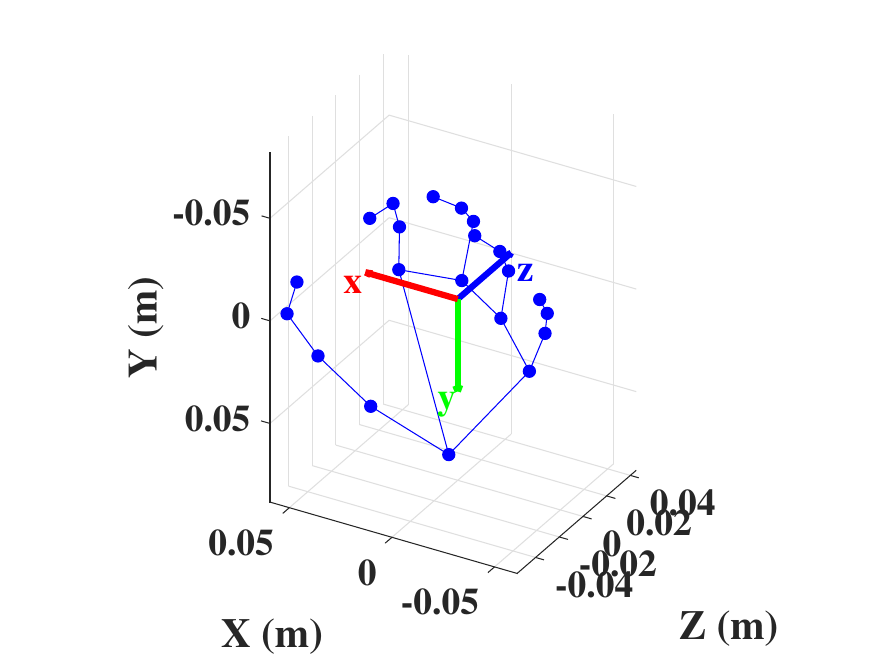}
    }
    \subfloat[Camera coordinate system]{
        \includegraphics[width=0.3\linewidth]{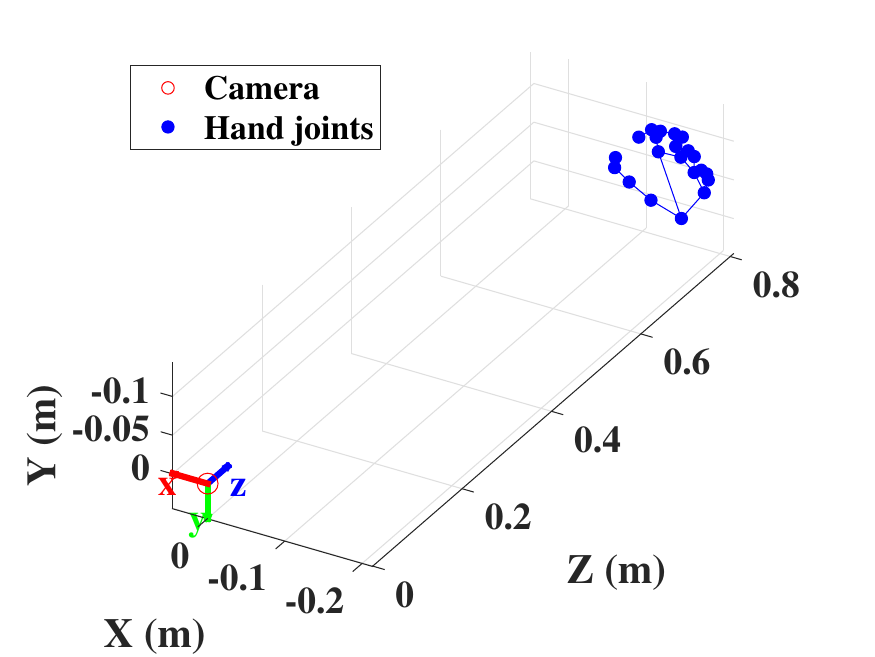}
    }
    \caption{Illustration of three coordinate systems.}
    \label{fig:perspective}
    % \vspace{-0.3cm}
\end{figure*}

For the elaboration convenience, we first introduce the following three coordinate systems. The two-dimensional (2D) pixel coordinate system in the unit of pixels is used to identify the positions of hand keypoints in each video frame. The origin of the pixel coordinate system is usually at the upper left corner of each frame, as shown in Fig. \ref{fig:perspective}. The three-dimensional (3D) hand world coordinate system in the unit of meters  measures the positions of hand keypoints in the real world with respect to the hand center. Moreover, the 3D camera coordinate system in the unit of meters  measures the positions of hand keypoints with respect to the static camera lens, which captures the videos. The \textit{Mediapipe} is able to identify the $21$ keypoints, localize them in the first two coordinate systems. Because the hand center is usually in motion and the camera is static, the trajectories in the camera coordinate system instead of in the hand world coordinate system, could be used to calculate the Doppler frequencies. Thus, the coordinates of hand keypoints $\mathbf{p}_{i}(t)$, $i=1,2,...,21$, transmitter $\mathbf{p}_\text{t}$ and receiver $\mathbf{p}_\text{r}$, defined in the previous section should be measured in the camera coordinate system. The above three coordinate systems are illustrated in Fig.~\ref{fig:perspective}, as referenced.

Define the coordinates of the $i$-th keypoint ($i=1,2,...,21$) in the pixel, hand world and camera coordinate systems as $(u_i,v_i)$, $(x_i^\text{w},y_i^\text{w},z_i^\text{w})$, and $(x_i,y_i,z_i)$, respectively, where the snapshot index $t$ is ignored in this section for the simplicity of elaboration. Let $f$ be the focal length in the unit of pixels, $(c_x,c_y)$ be the coordinates of image center in the pixel coordinate system, we define the camera intrinsic matrix $\mathbf A$ as
\begin{align}
    \mathbf{A} = \begin{bmatrix}
    f & 0 & c_x\\
    0 & f & c_y\\
    0 & 0 & 1
    \end{bmatrix}.
    \label{eq:intrisic}
\end{align}
Hence, the relation between the 2D pixel and 3D camera coordinate systems can be expressed as 
\begin{align}
    % between camera and 2d
    z_i[u_i \ v_i \ 1]^T  = \mathbf{A} [x_i \ y_i \ z_i]^T.
\end{align}

Let $\mathbf{R}\in\mathbb{R}^{3 \times 3}$ and $\mathbf{t}$ be the rotation matrix and translation vector from hand world coordinate system to camera coordinate system, we define the camera extrinsic matrix $\mathbf{T}$ and perspective projection matrix $\mathbf{\Pi}$ as follows:
\begin{align}
    \mathbf{T} = \begin{bmatrix}
    \mathbf{R} & \mathbf{t}\\
    \mathbf{0}_{1 \times 3} & 1
    \end{bmatrix},
    \label{eq:extrinsic}
\end{align}
\begin{align}
    \mathbf{\Pi} = 
    \begin{bmatrix}
    \mathbf{I}_{3\times 3}  & \mathbf{1}_{3\times 1}
    \end{bmatrix},
\end{align}
where $\mathbf{I}_{3\times 3}$ denotes a $3\times3$ identity matrix, $\mathbf{0}_{1 \times 3}$ and $\mathbf{1}_{3 \times 1}$ are the three-dimensional row and column vectors with all 0 and 1 entries respectively. According to \cite{marchand:hal-01246370}, the relations between the hand world and camera coordinate systems are given by 
\begin{align}
    % between camera and hand world
    [x_i \ y_i \ z_i \ 1]^T = \textbf{T} [x_i^\text{w} \ y_i^\text{w} \ z_i^\text{w} \ 1]^T.
    \label{eq:w2c_transformation}
\end{align}

As a result, the relation between the hand world and the pixel coordinate system could be described as
\begin{align}
    & z_i[u_i \ v_i \ 1]^T 
    =\mathbf{A} \mathbf{\Pi} \mathbf{T} [ x_i^\text{w} \ y_i^\text{w} \ z_i^\text{w} \ 1 ]^T.
    \label{eq:pnp}
\end{align}
For the elaboration convenience, we denote the projection from the hand world coordinate system to the pixel coordinate system as the following function $\mathcal{P}$: 
\begin{align} \label{eqn:projection}
    [u_i \ v_i]^T &= \mathcal{P}([ x_i^\text{w} \ y_i^\text{w} \ z_i^\text{w}]^T,\mathbf{R}, \mathbf{t}, \mathbf{A}) \nonumber \\
    &= \frac{1}{z_{i}}[\mathbf{I}_{2 \times 2} \ \mathbf{0}_{2 \times 1}]\mathbf{A}\underbrace{(\mathbf{R}[ x_i^\text{w} \ y_i^\text{w} \ z_i^\text{w}]^T+\mathbf{t})}_{=[x_i \ y_i \ z_i]^T} .
\end{align}

The \textit{Mediapipe} could provide the coordinates $(u_i,v_i)$ and $(x_i^\text{w},y_i^\text{w},z_i^\text{w})$ of all the keypoints ($i=1,2,...,21$) in each video frame. Hence, their coordinates in the camera coordinate system can be calculated with the knowledge of the rotation matrix $\mathbf{R}$ and translation vector $\mathbf{t}$. 

In fact, the parameters in the camera intrinsic matrix $\mathbf{A}$ can be measured in advance, the rotation matrix $\mathbf{R}$ and translation vector $\mathbf{t}$ can be estimated via (\ref{eqn:projection}) for $i=1,2,...,21$. Particularly, given the coordinates of the $21$ keypoints in the pixel and hand world coordinate systems, the detection of the rotation matrix $\mathbf{R}$ and translation vector $\mathbf{t}$ can be formulated as follows. 
\begin{align}
\mathop{\mathrm{min}}_{\mathbf{R},\mathbf{t}} \quad&
\sum_{i=1}^{21} | (u_i, v_i) - \mathcal{P}([ x_i^\text{w} \ y_i^\text{w} \ z_i^\text{w}]^T,\mathbf{R}, \mathbf{t}, \mathbf{A}) |^2, \nonumber \\
\mathrm{s.t.} \quad& \mathbf{R}(\mathbf{R})^T = \mathbf{I}_{3 \times 3},\ \mathrm{det}(\mathbf{R})=1,
\label{eq:pnp_min}
\end{align}
where $\mathrm{det}(.)$ represents the determinant of a matrix.

The above problem is referred to as the Perspective-n-Point (PnP) problem \cite{marchand:hal-01246370}. It can be solved via the \textit{cv2.solvePnP} function from the popular computer vision library \textit{OpenCV} \cite{opencv_library}, where the Levenberg-Marquardt optimization method \cite{levenberg1944method} is adopted.

\subsection{Motion Smoothing and Snapshot Interpolation}
\label{sec:smoothing and interpolation}
\begin{figure}[t]
    \centering
    \subfloat[]{
        \includegraphics[width=0.45\linewidth]{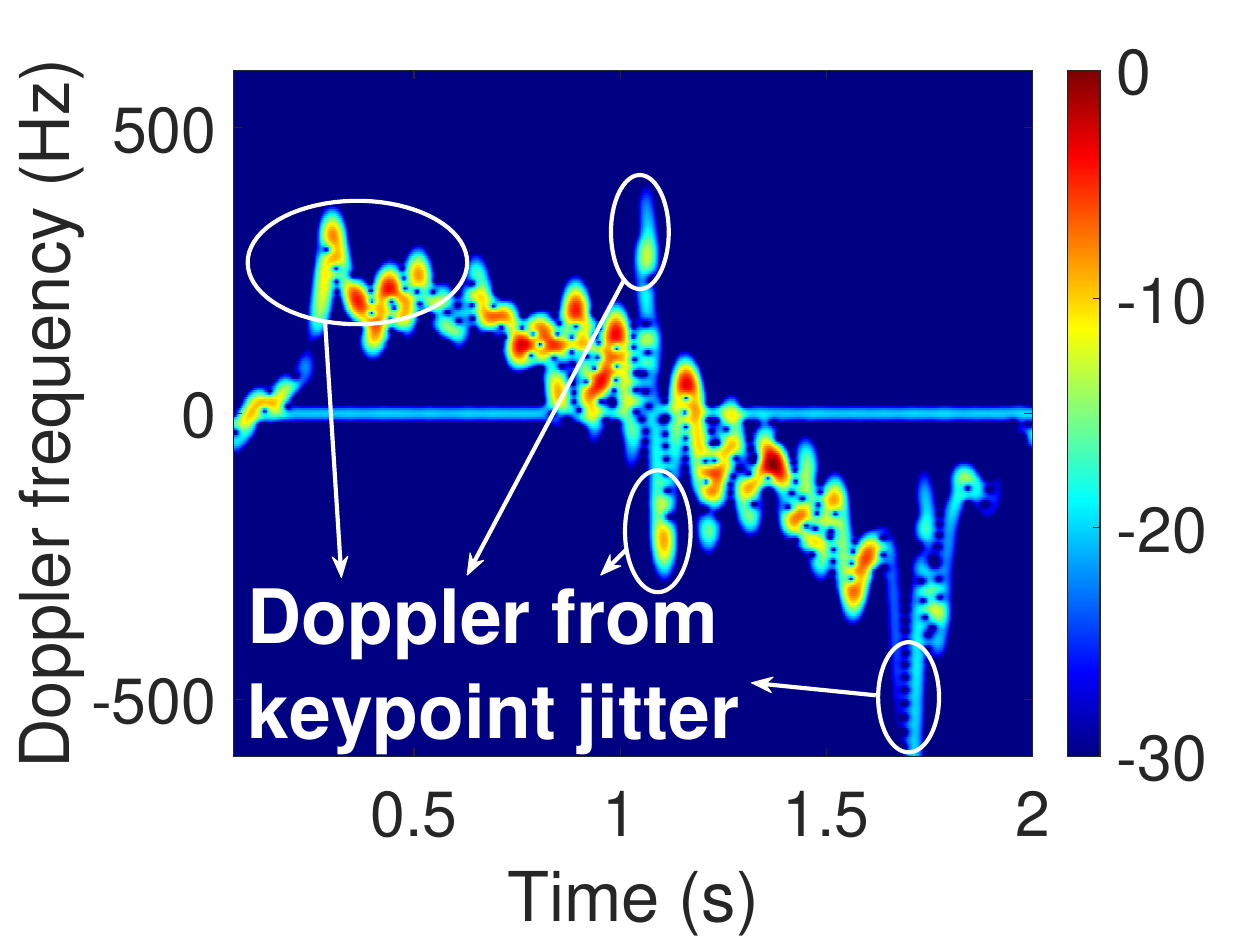}
    }
    \subfloat[]{
        \includegraphics[width=0.45\linewidth]{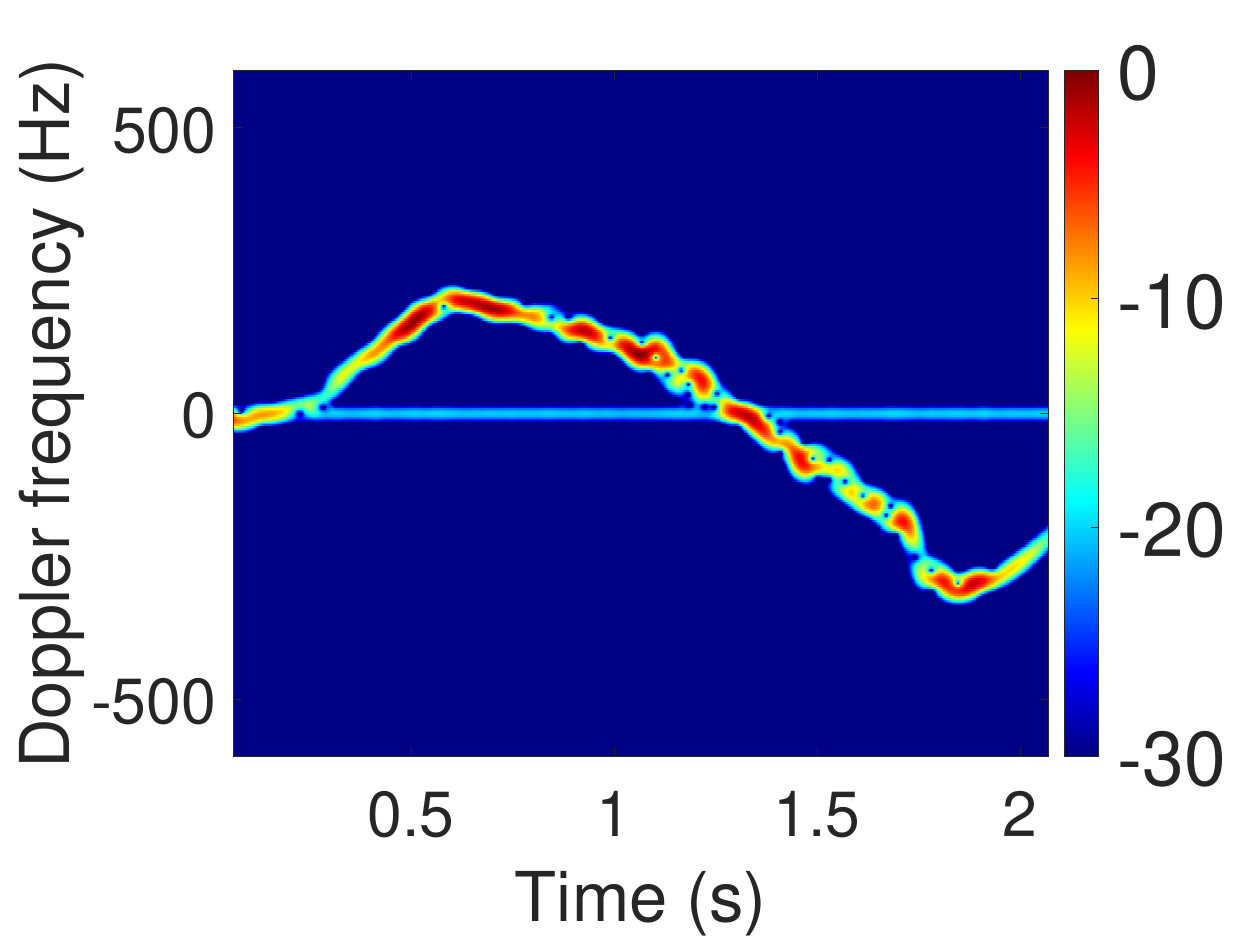}
    }
    \caption{Comparison of simulated spectrograms via {\SystemName}. (a) before one-euro filter smoothing; (b) after  one-euro filter smoothing.}
    \label{fig:smoothing}
    % \vspace{-0.3cm}
\end{figure}
Because of the errors of keypoint detection with \textit{Mediapipe}, there might be fake hops or jitters in the detected trajectories of keypoints, which do not exist actually. This will lead to the false alarm of high Doppler frequencies (as depicted in Fig.~\ref{fig:smoothing}). In order to generate a high-fidelity dataset for gesture recognition model training, a low-pass filter, namely one-euro filter \cite{casiez20121}, is proposed to smooth both trajectories and velocities, followed by snapshot interpolation between neighboring video frames.  

Let $\mathbf{q}_{i,k}=[x_{i,k}\ y_{i,k} \ z_{i,k}]^T$ and $\hat{\mathbf q}_{i,k}=[{\hat x}_{i,k}\ \hat{y}_{i,k} \ \hat{z}_{i,k}]^T$ be the positions of the $i$-th keypoint in the $k$-th frame before and after the low-pass filtering respectively, $\Dot{\mathbf{q}}_{i,k}=[\Dot{x}_{i,k}\ \Dot{y}_{i,k} \ \Dot{z}_{i,k}]^T$ and $\hat{\Dot{\mathbf{q}}}_{i,k}=[\hat{\Dot{x}}_{i,k}\ \hat{\Dot{y}}_{i,k} \ \hat{\Dot{z}}_{i,k}]^T$ be the estimated velocities of the $i$-th keypoint in the $k$-th frame before and after the low-pass filtering respectively. Initializing $\hat{\mathbf q}_{i,1}$ with $\mathbf{q}_{i,1}$, the trajectory smoothing for the $i$-th keypoint in the $k$-th frame is given by
\begin{align}
    \hat{o}_{i,k} = \alpha_{i,k} o_{i,k} + (1-\alpha_{i,k})\hat{o}_{i,k-1},\quad \forall i,k \geq 2
    \label{eq:lpf_position}
\end{align}
where the notation $o$ represents the dimensions of $x$, $y$ and $z$, respectively, and
\begin{align}
    \alpha_{i,k}  = \frac{1}{1+\frac{1}{2\pi {\Delta t}_\text{v}(f_{\text{c}_\text{min}} + \beta |\hat{\Dot{o}}_{i,k}|)}} \nonumber
\end{align}
is the smoothing factor, ${\Delta t}_\text{v}$ is the video frame interval, $f_{\text{c}_\text{min}}$ is the minimum cutoff frequency, $\beta$ is the speed coefficient of update. Moreover, the velocity in the above equation can be calculated as
\begin{align}
    \hat{\Dot{o}}_{i,k} = \gamma \Dot{o}_{i,k} + (1 - \gamma)\hat{\Dot{o}}_{i,k-1},\quad \forall i,k \geq 2
    \label{eq:lpf_velocity}
\end{align}
where $\Dot{o}_{i,k}= (o_{i,k}-\hat{o}_{i,k-1})/{\Delta t}_\text{v}$, $\hat{\Dot o}_{i,1}$ is initialized with $0$, $\gamma$ is the fixed smoothing factor.

\begin{algorithm}[htb]
    \caption{One-euro low-path filter for keypoint trajectory smoothing.}
    \label{alg:one-euro}
    \begin{algorithmic}[1]
        \State \textbf{Input:} 
        \begin{itemize}
            \item $\{\mathbf{q}_{i,k}=[x_{i,k}\ y_{i,k} \ z_{i,k}]^T | i \in \{1,\ldots,21\},k \in\{1,\ldots,K\}\}$, where $\mathbf{q}_{i,k}$ denotes the location of the $i$-th keypoint in the $k$-th frame.
            \item ${f_\text{c}}_\text{min}$: Minimum cutoff frequency for position. 
            \item $\beta$: Speed coefficient. 
            \item $\gamma$: Smoothing factor for velocity.
            \item $\Delta t_\text{v}$: Video frame interval.
        \end{itemize} 
        \State \textbf{Output:} 
        \begin{itemize}
            \item $\{\mathbf{\hat{q}}_{i,k}=[\hat{x}_{i,k}\ \hat{y}_{i,k} \ \hat{z}_{i,k}]^T| i \in \{1,\ldots,21\},k \in\{1,\ldots,K\}\}$: where $\mathbf{\hat{q}}_{i,k}$ denotes the location of the $i$-th keypoint in the $k$-th frame after smoothing.
        \end{itemize}
        \For{$k \gets 2$ to $K$} \Comment{Iteration over frames.}
            \For{$i \gets 1$ to $21$} \Comment{Iteration over keypoints.}
                \For{$o$ represents the dimensions of $x$, $y$, and $z$ respectively}  
                    \State $\hat{o}_{i,1} \gets o_{i,1}$, $\hat{\Dot{o}}_{i,1} \gets 0$ 
                    \State $\Dot{o}_{i,k}= (o_{i,k}-\hat{o}_{i,k-1})/{\Delta t}_\text{v}$
                    \State $\hat{\Dot{o}}_{i,k} = \gamma \Dot{o}_{i,k} + (1 - \gamma)\hat{\Dot{o}}_{i,k-1}$ \Comment{Equation (21): smooth velocity.}
                    \State $\alpha_{i,k}  = \frac{1}{1+\frac{1}{2\pi {\Delta t}_\text{v}(f_{\text{c}_\text{min}} + \beta |\hat{\Dot{o}}_{i,k}|)}}$ \Comment{Update smoothing factor for position.}
                    \State $\hat{o}_{i,k} = \alpha_{i,k} o_{i,k} + (1-\alpha_{i,k})\hat{o}_{i,k-1}$ \Comment{Equation (20): smooth position.}
                \EndFor
                \State $\mathbf{\hat{q}}_{i,k}=[\hat{x}_{i,k}\ \hat{y}_{i,k} \ \hat{z}_{i,k}]^T$
            \EndFor
        \EndFor
        \State \textbf{return} $\{\mathbf{\hat{q}}_{i,k}=[\hat{x}_{i,k}\ \hat{y}_{i,k} \ \hat{z}_{i,k}]^T| i \in \{1,\ldots,21\},k \in\{1,\ldots,K\}\}$
    \end{algorithmic}
\end{algorithm}

The overall smoothing procedure via one-euro filter is illustrated in Alg.~\ref{alg:one-euro}. In fact, the smoothing of the $i$-th keypoint's velocity $\hat{\Dot{o}}_{i,k}$ and trajectory $\hat{o}_{i,k}$ in the $k$-th frame is conducted by repeating two first-order low-pass filters \eqref{eq:lpf_velocity} and \eqref{eq:lpf_position} to the position and velocity of the $i$-th keypoint. This procedure effectively eliminates false hops or jitters in the detected keypoint trajectories while preserving the motion features. An example of the smoothing result is shown in Fig.~\ref{fig:smoothing}.

Finally, we adopt the cubic spline interpolation method \cite{de1978practical} to insert $\Delta t_\text{v}/ \Delta t_\text{s} - 1$ positions of the $i$-th keypoint ($\forall i$) between every two neighboring frames (say $\hat{\mathbf{q}}_{i,k}$ and $\hat{\mathbf{q}}_{i,k+1}$, $\forall k$), and denote the position of the $i$-th keypoint in the $t$-th snapshot as $\mathbf{p}_i(t)$.

\section{Evaluation of {\SystemName} Simulator}
\label{sec:result}

\begin{figure*}[t]
    \centering
    \includegraphics[width = \linewidth]{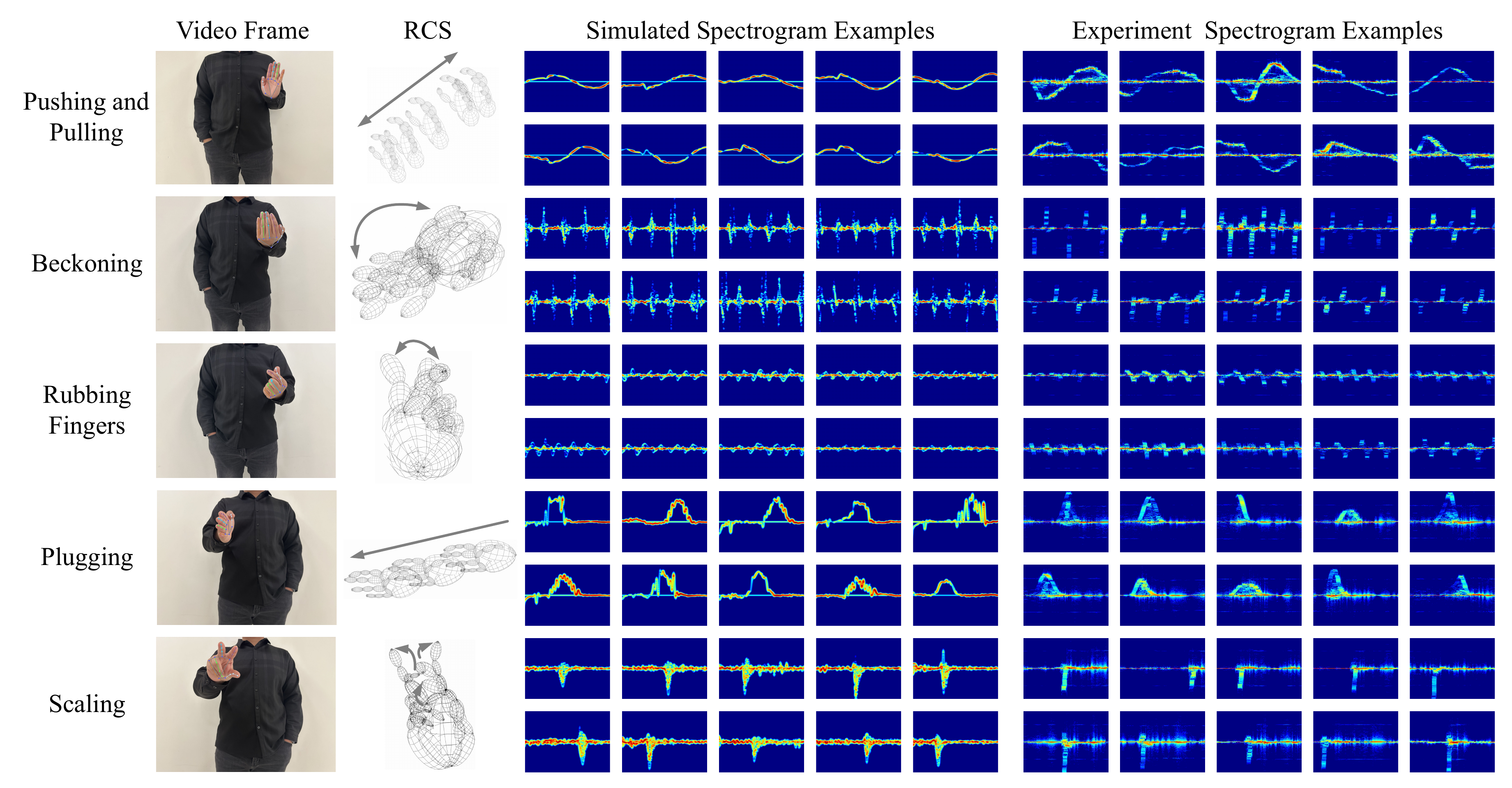}
    \caption{Illustration of the simulated and experimental dataset from {\SystemName}, where some examples of spectrogram are plotted.}
    \label{fig:dataset}
\end{figure*}

\begin{figure}[t]
    \centering
    \includegraphics[width = \linewidth]{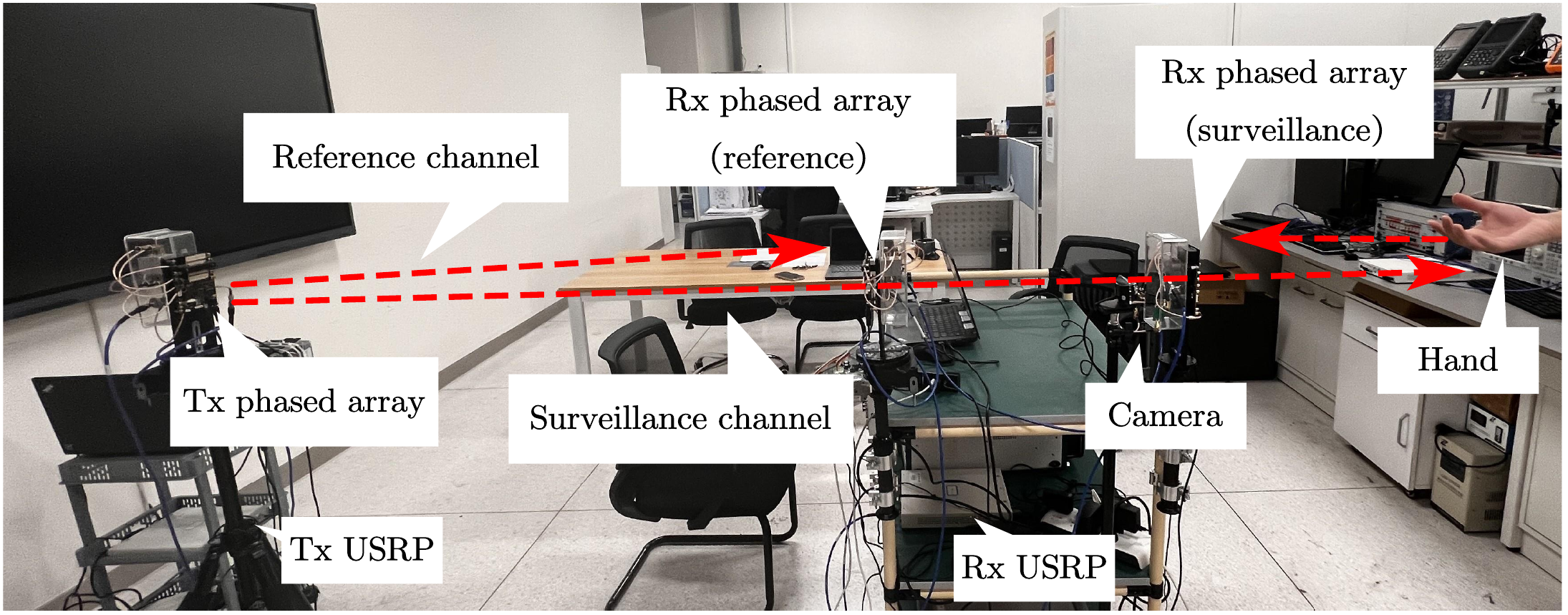}
    \caption{Facilities and scenario of experiment.}
    \label{fig:exp}
    % \vspace{-0.3cm}
\end{figure}

In this section, the high fidelity of the {\SystemName} simulator in the applications of gesture recognition is demonstrated. Specifically, the generation of gesture datasets via  {\SystemName} simulator and real measurement is first elaborated. Then, the recognition performance via the above two datasets is discussed.

\subsection{Simulation and Experimental Datasets}
In order to verify the quality of the dataset generated by {\SystemName} simulator, $500$ clips of videos on $5$ gestures, including ``Pushing and Pulling'', ``Beckoning'', ``Rubbing Fingers'', ``Plugging'' (slicing forward with all fingers together), and ``Scaling'' (spreading thumb, index finger, middle finger) were recorded using a normal monocular camera at a rate of $30$ frames per second (fps). The motion data for hand model is then extracted via the video gesture catcher.

On the other hand, in the channel generator, the locations of transmitter, receiver and target hand center are $[0m, -0.1m, -1.5m]$, $[0.2m, -0.1m, 0.1m]$, and $[0m, 0m, 0.4\sim0.8m]$, respectively.
Moreover, in order to model the target-unrelated channel, $K$ static RCSs are randomly generated from a normal distribution with a mean value of $0.005\, m^2$ and a standard deviation of $0.001\, m^2$. These RCSs are associated with scatterers that are randomly located within a $2\, m \times 2\, m \times 2\, m$ cuboid centered at the receiver. The positions of these scatterers are used to calculate the associated parameters $G_\text{t}^k$, $G_\text{r}^k$, $R_\text{t}^k$, and $R_\text{r}^k$.

Thus, $100$ sequences of channel impulse responses for each gesture are obtained via the proposed {\SystemName} simulator with a sampling rate of $2000$ snapshots per second. Then, one spectrogram, illustrating the Doppler frequency versus time, is calculated for each video clip (each sequence of channel impulse responses) by applying the short-time Fourier transform (STFT) with a window of $0.125$ seconds ($250$ snapshots). 
As a result, a simulated dataset of \blue{$500$} spectrograms for the recognition of \blue{5} gestures is obtained as illustrated in Fig.~\ref{fig:dataset}.

In order to measure the real Doppler spectrum of gestures, an integrated passive sensing and communication system working on millimeter wave (mmWave) band is developed as in our previous work \cite{li2022passive}. As illustrated in Fig.~\ref{fig:exp}, at the transmitter, an NI USRP-2954R \cite{USRP2954} is utilized to generate an intermediate frequency (IF) signal at $500$ MHz, which is subsequently up-converted to 60 GHz and transmitted using a Sivers $60$ GHz phased array\cite{EVK06002}. At the receiver, two phased arrays are connected to a single USRP device to receive signals from the reference and surveillance channels, respectively. The transmit mmWave signal is modulated via orthogonal frequency-division multiplexing (OFDM). The carrier frequency is $60.48$ GHz and the signal bandwidth is $5$ MHz. 

\newcommand{\widthSpectrogram}{0.18\linewidth}
\begin{figure*}[htb]
    \centering
    \subfloat[Pushing \& Pulling (real)]{
        \includegraphics[width = \widthSpectrogram]{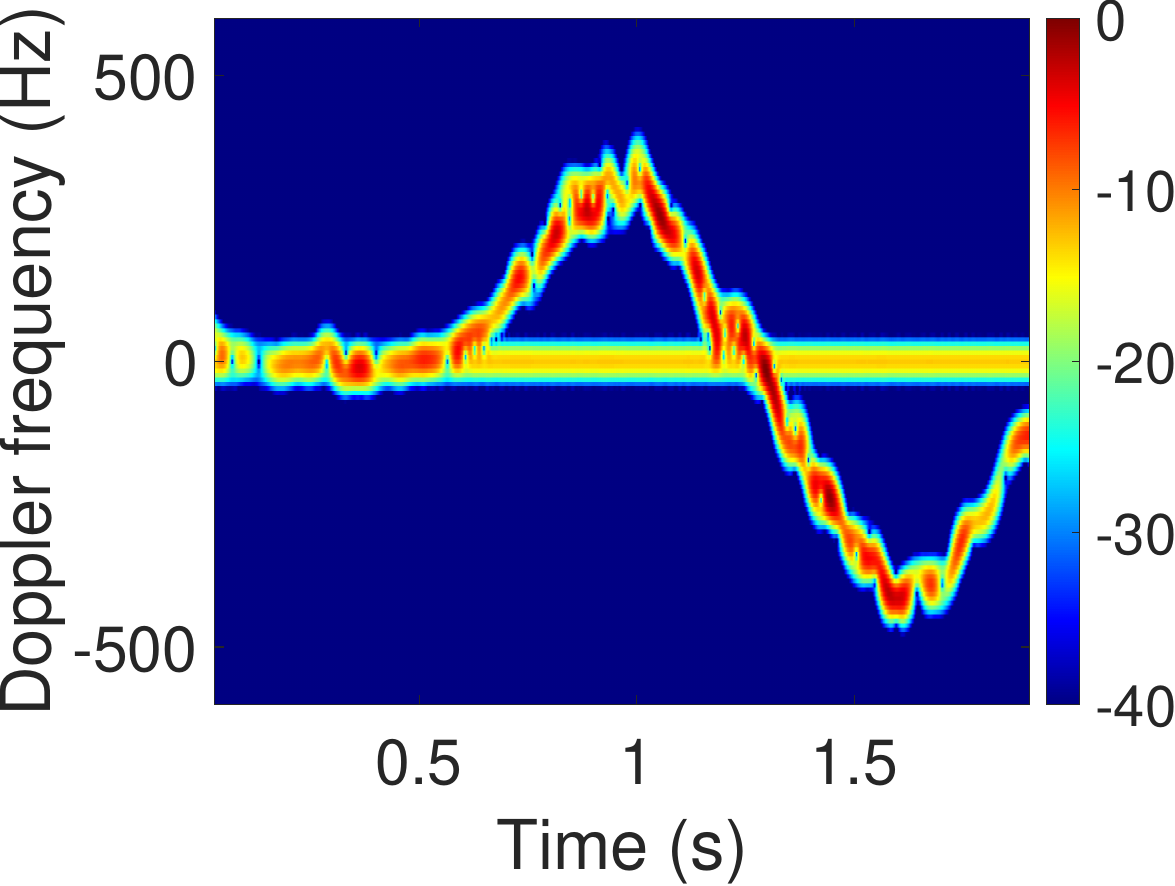}
    }
    \subfloat[Beckoning (real)]{
        \includegraphics[width = \widthSpectrogram]{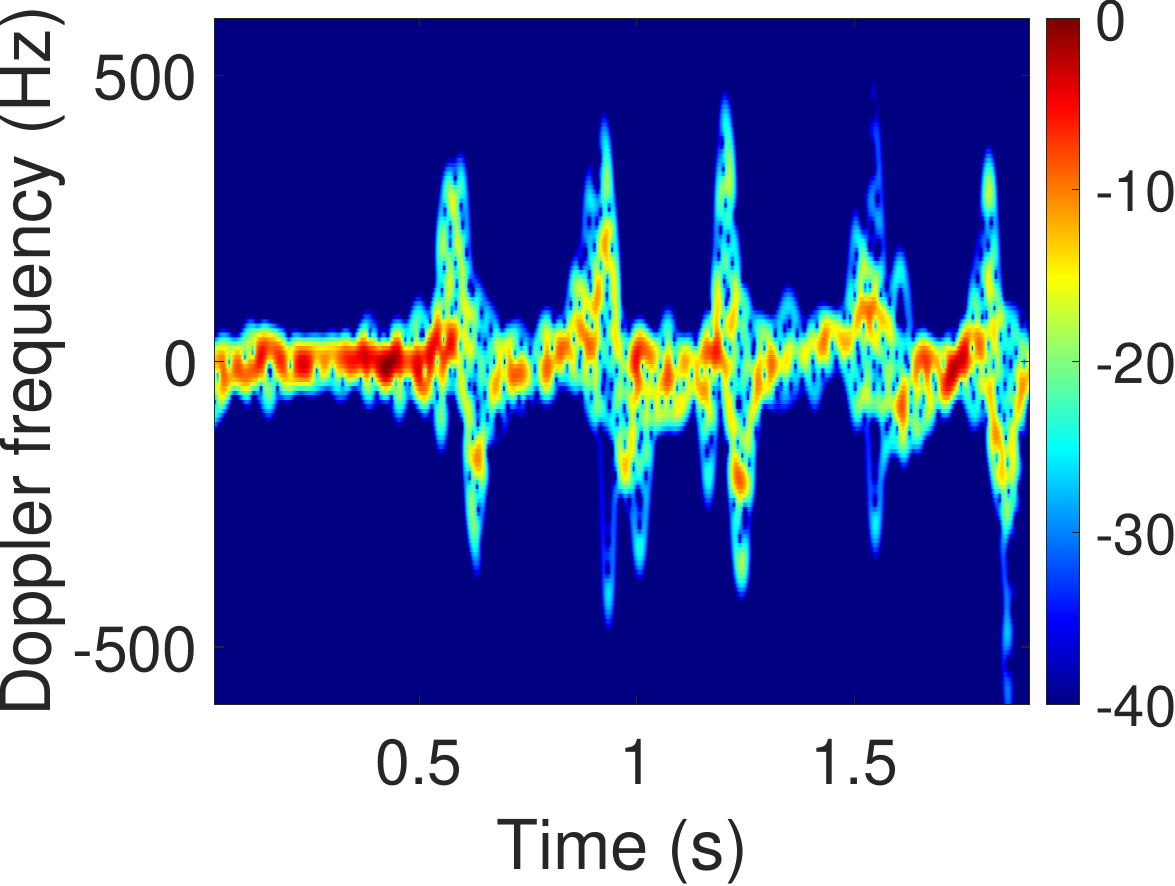}
    }
    \subfloat[Rubbing Fingers (real)]{
        \includegraphics[width = \widthSpectrogram]{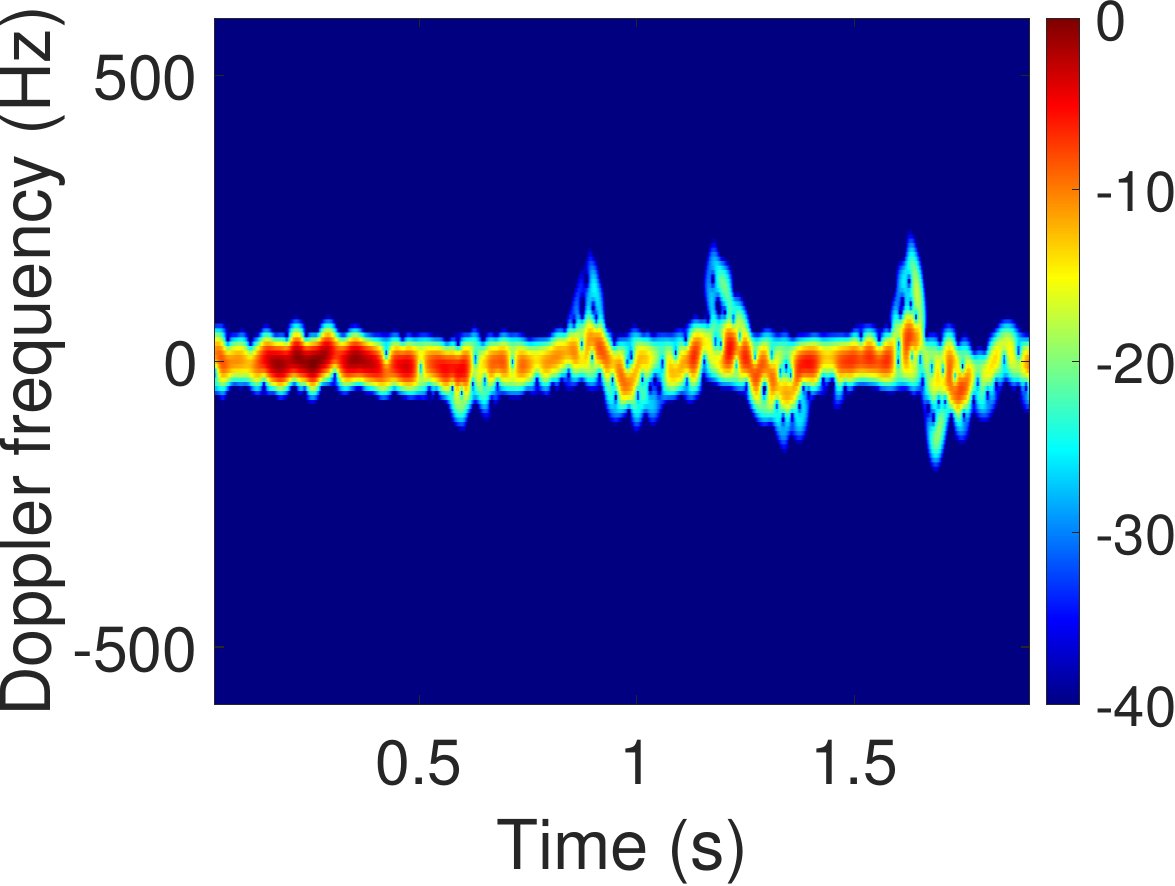}
    }
    \subfloat[Plugging (real)]{
        \includegraphics[width = \widthSpectrogram]{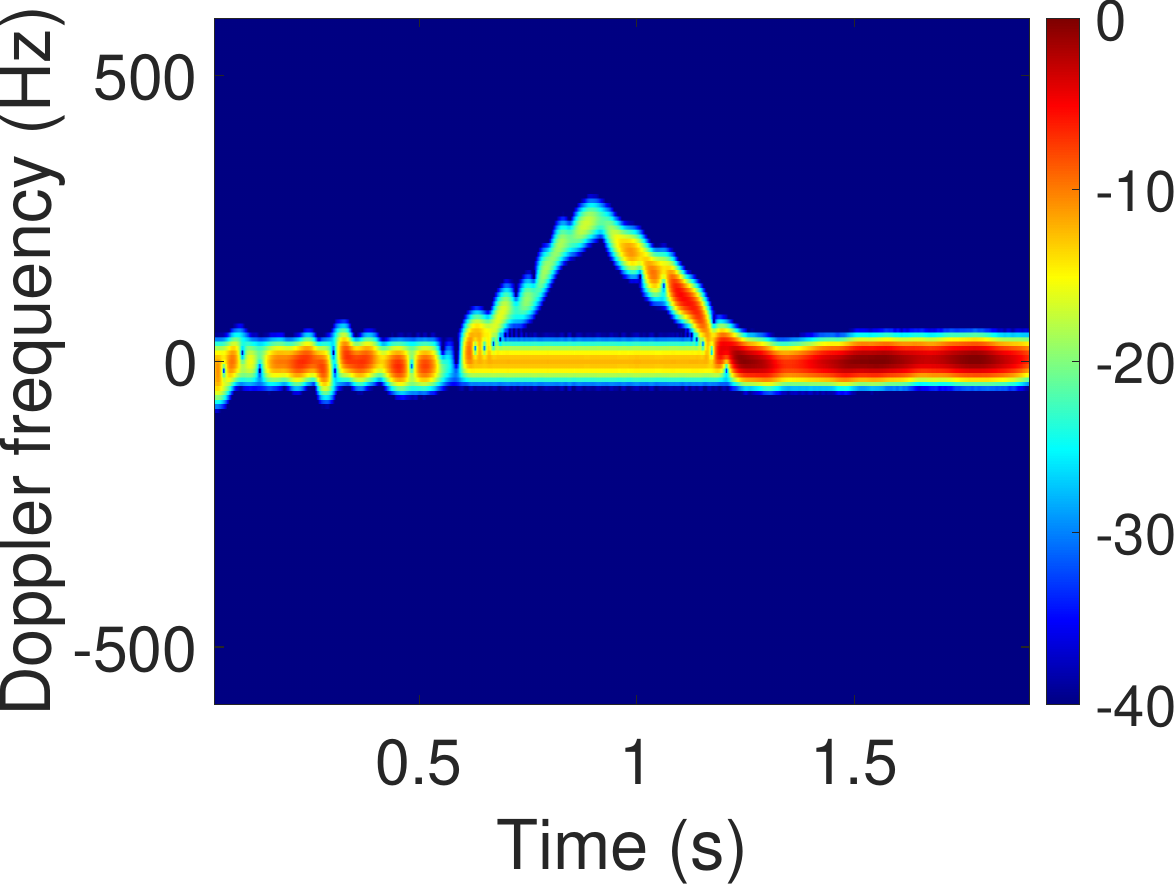}
    }
    \subfloat[Scaling (real)]{
        \includegraphics[width = \widthSpectrogram]{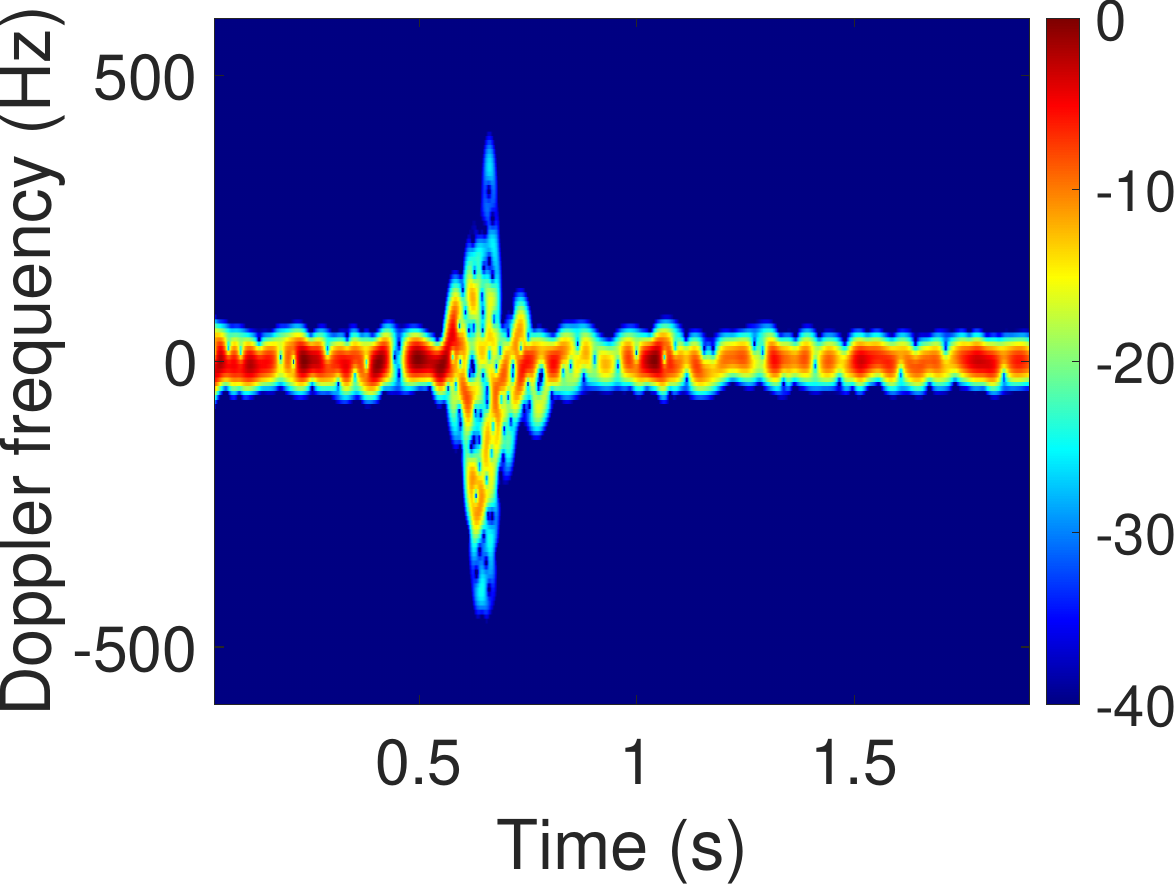}
    }
    \\  
    \subfloat[Pushing \& Pulling (sim)]{
        \includegraphics[width=\widthSpectrogram]{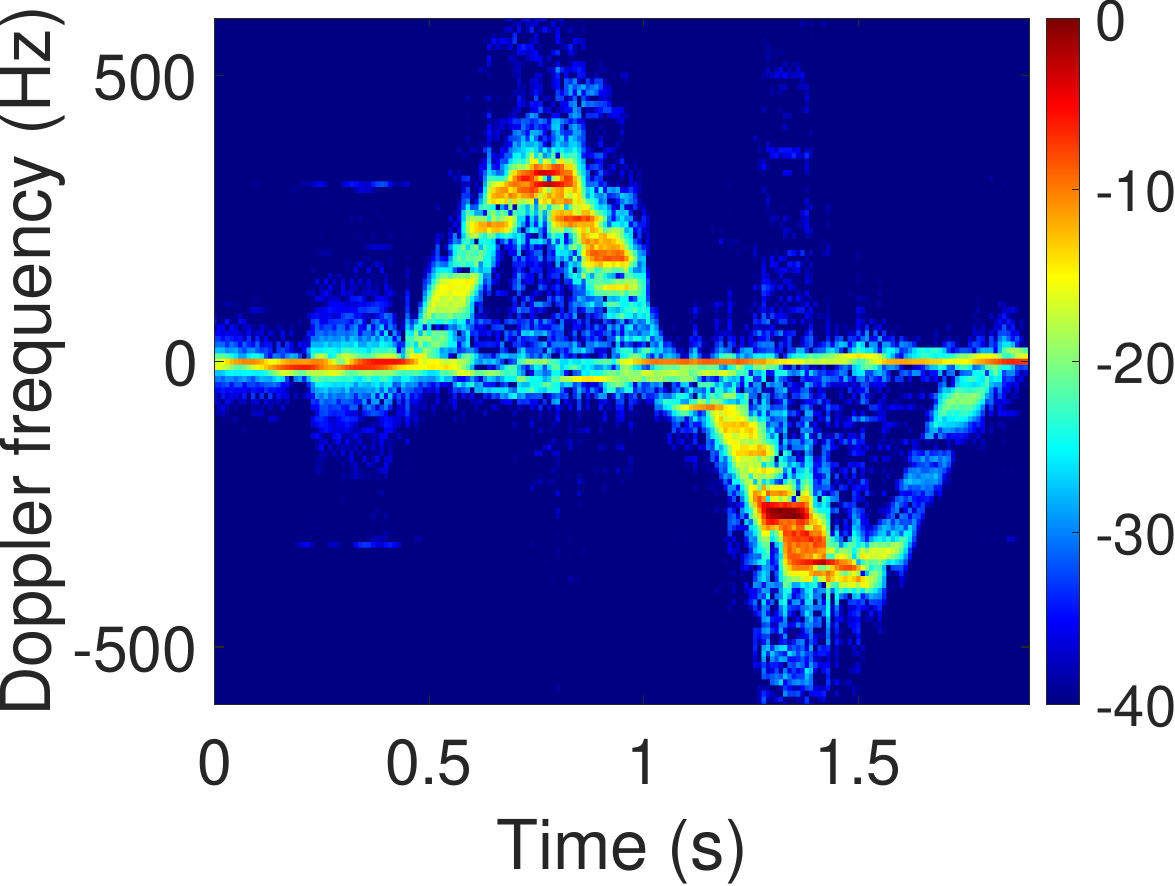}
    }
    \subfloat[Beckoning (sim)]{
        \includegraphics[width=\widthSpectrogram]{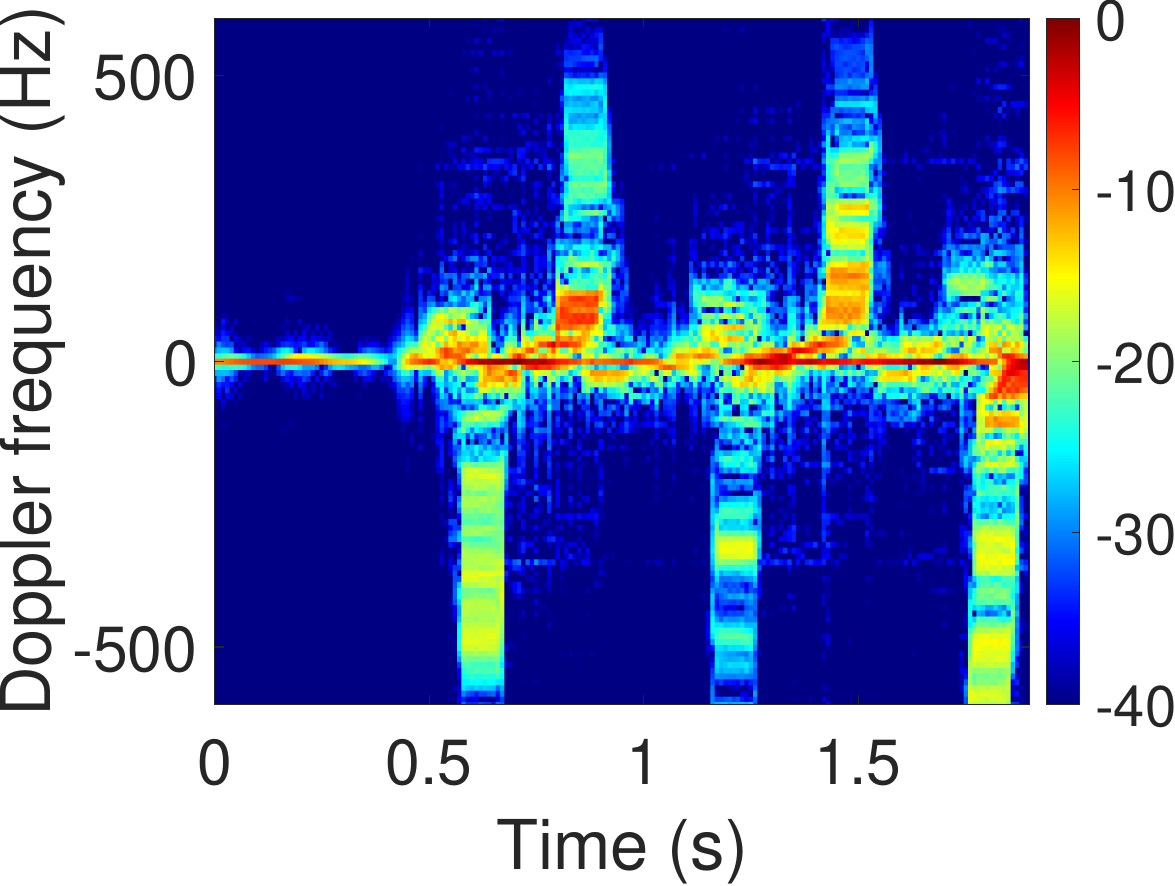}
    }
    \subfloat[Rubbing Fingers (sim)]{
        \includegraphics[width=\widthSpectrogram]{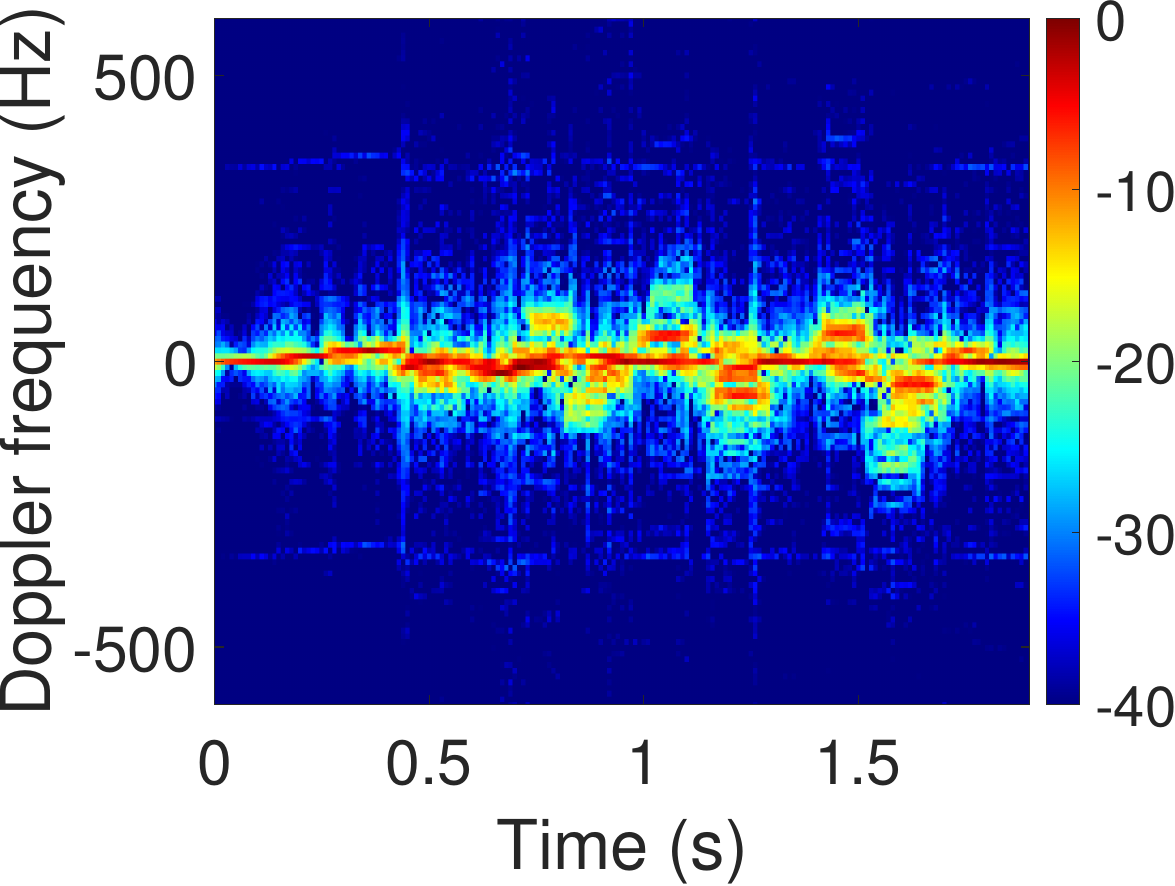
    }
    }
    \subfloat[Plugging (sim)]{
        \includegraphics[width=\widthSpectrogram]{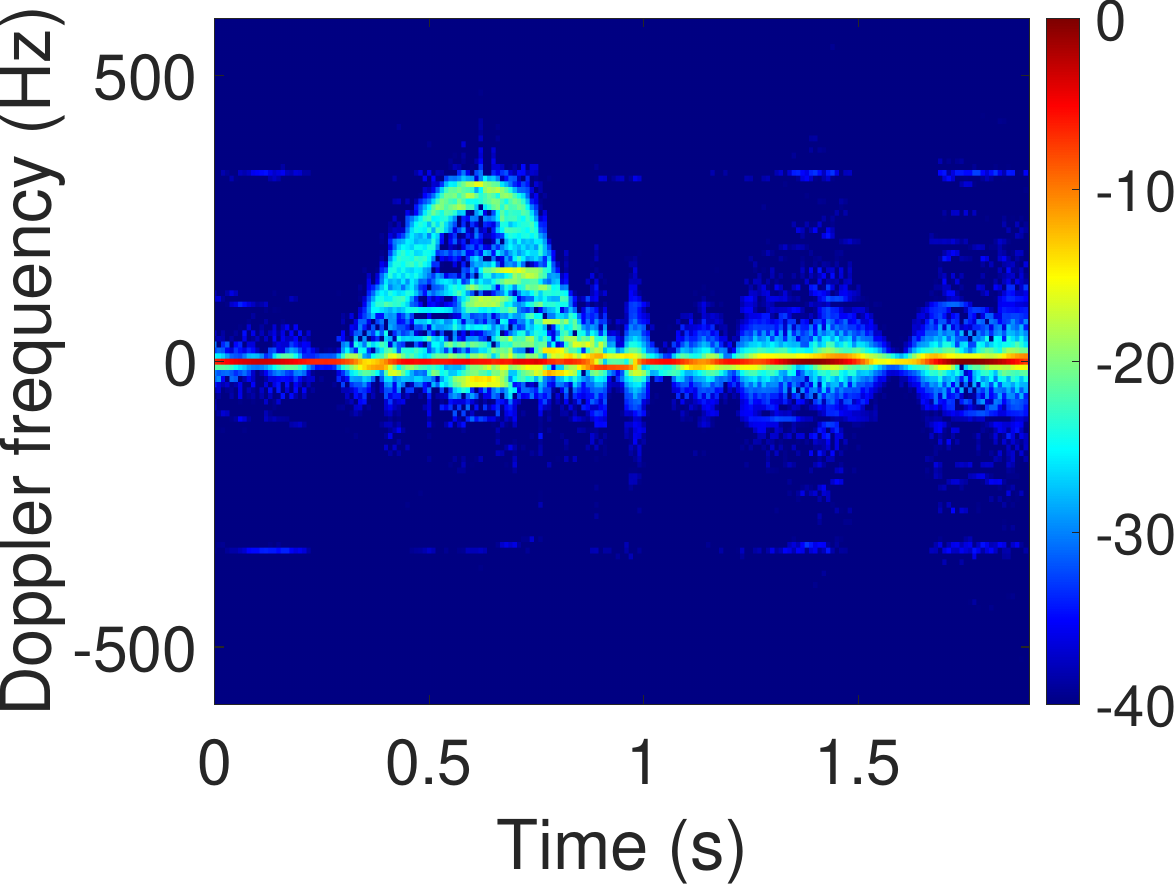}
    }
    \subfloat[Scaling (sim)]{
        \includegraphics[width=\widthSpectrogram]{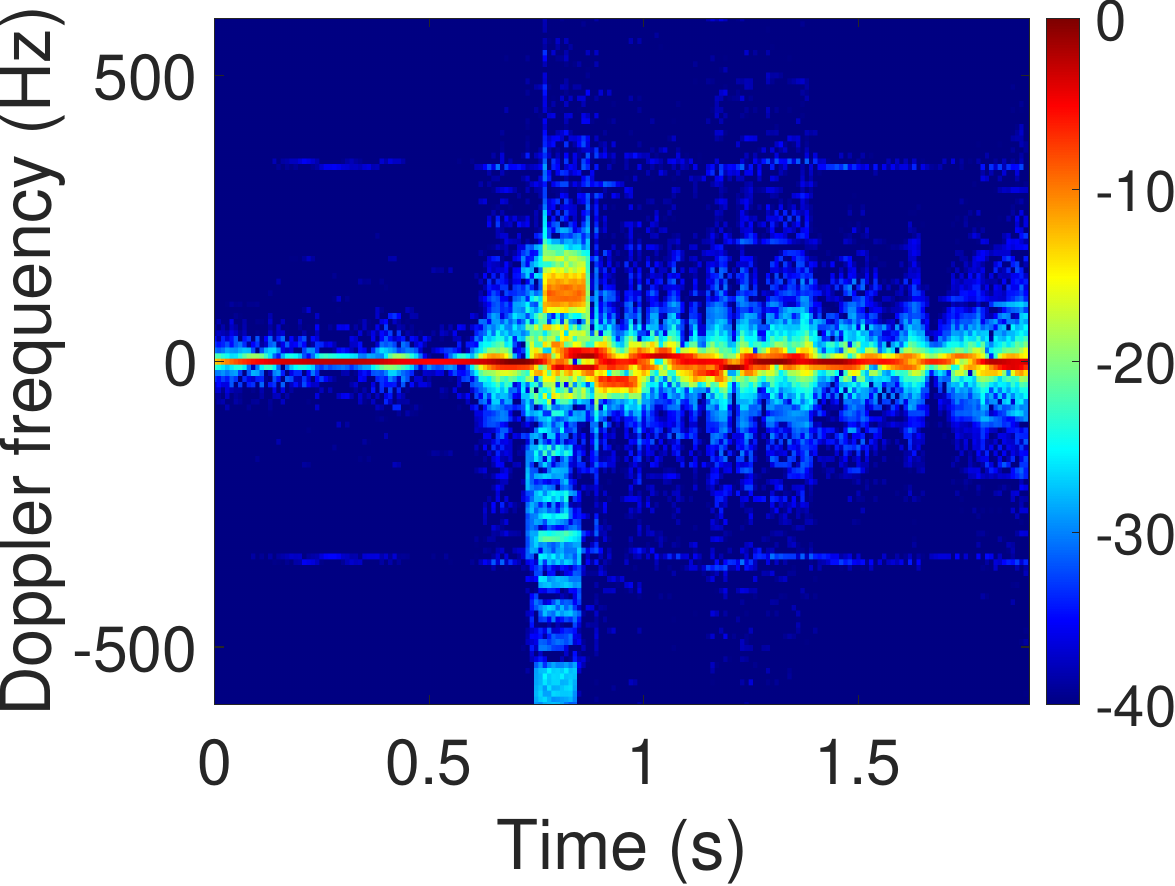}
    }  
    \caption{Spectrogram comparison of $5$ gestures generated by {\SystemName} (first row) and the experiment (second row).}
    \label{fig:exp-simu}
\end{figure*}  

In the experiment, the locations of the transmitter and receiver are consistent with those in the simulator. $100$ trials are measured for each gesture via the passive sensing system. Following the signal processing in \cite{li2022passive}, the spectrogram of hand gestures can be computed through the cross-ambiguity function (CAF). As a result, an experimental dataset with $100$ spectrograms per gesture is obtained, as illustrated in Fig.~\ref{fig:dataset}.

\begin{figure}[t]
    \centering
    \includegraphics[width=\linewidth]{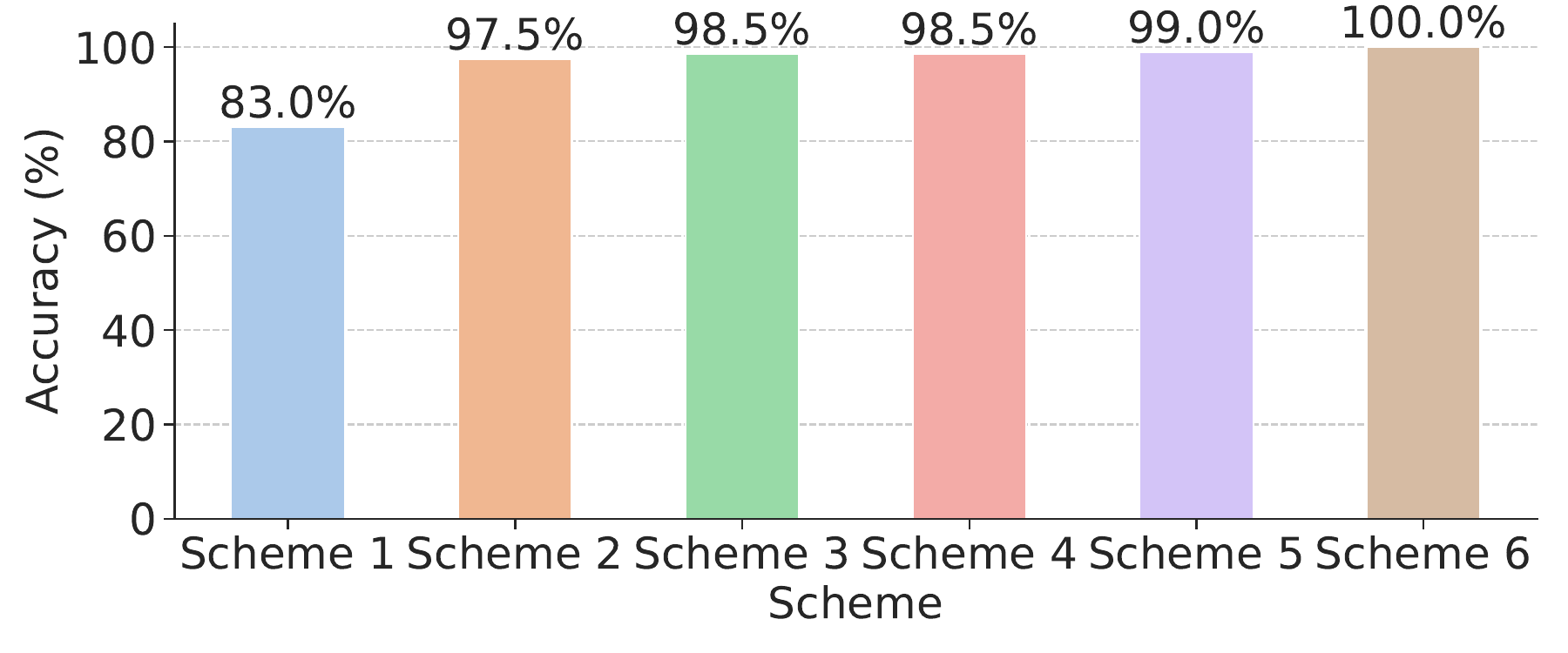} 
    \caption{Gesture recognition accuracy of the $6$ training and testing schemes.}
    \label{fig:training_result}
\end{figure}

\newcommand{\widthConfMatrix}{0.3\linewidth}
\begin{figure*}[t]
    \centering
    \subfloat[Scheme $1$]{
        \includegraphics[width=\widthConfMatrix]{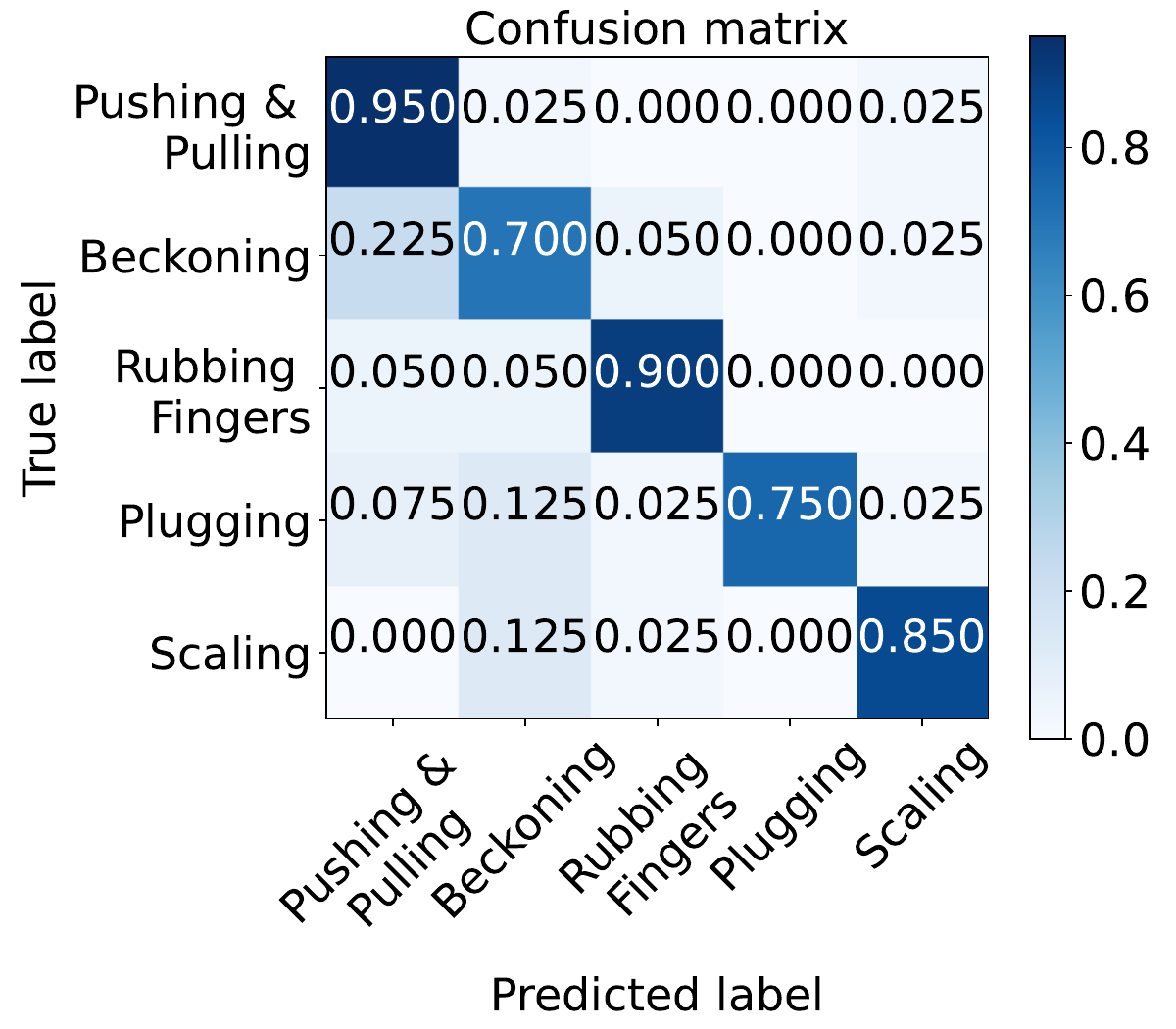}
        \label{fig:confusion_charts(a)}
    }
    \subfloat[Scheme $2$]{
        \includegraphics[width=\widthConfMatrix]{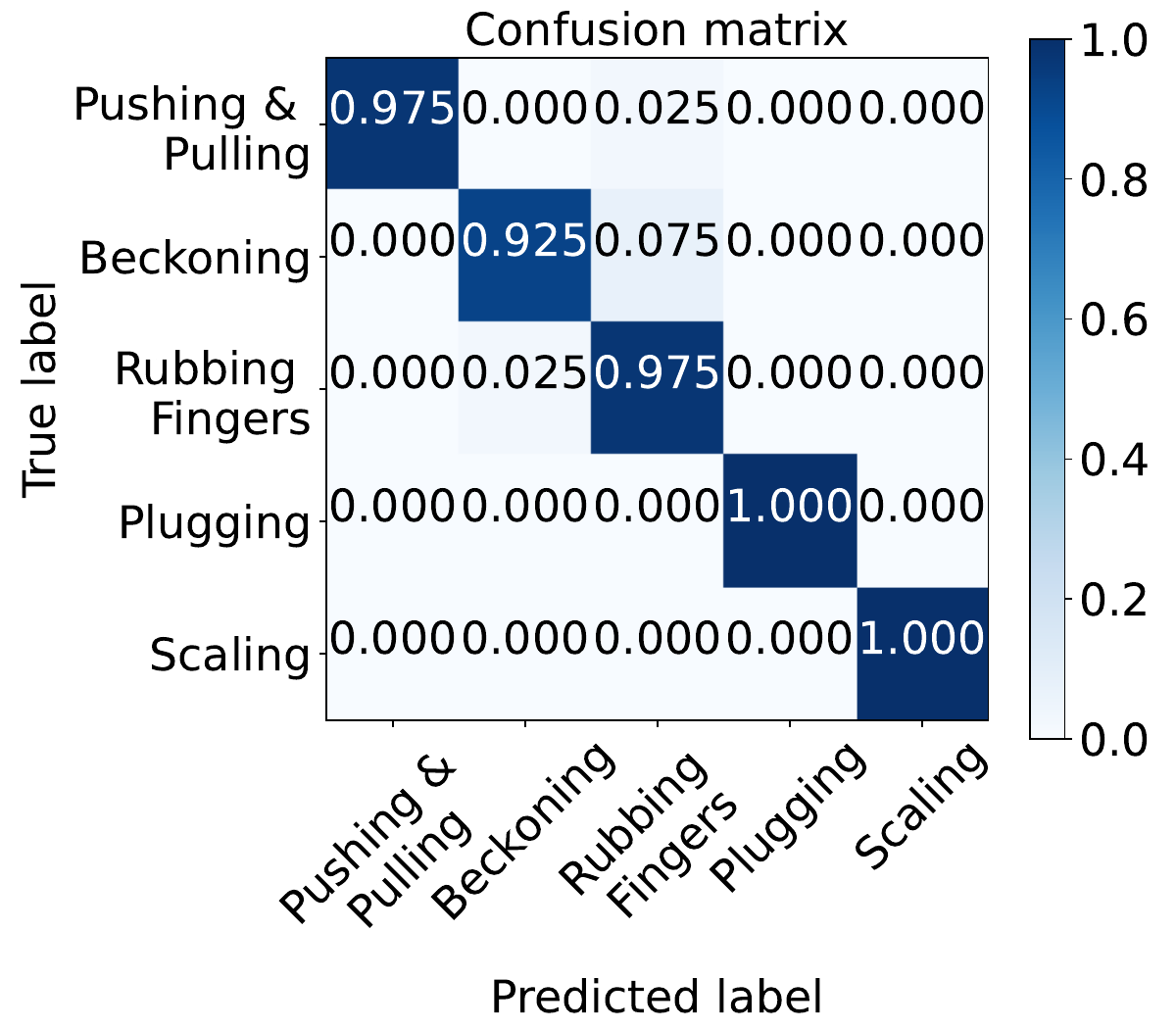}
    }
    \subfloat[Scheme $3$]{
        \includegraphics[width=\widthConfMatrix]{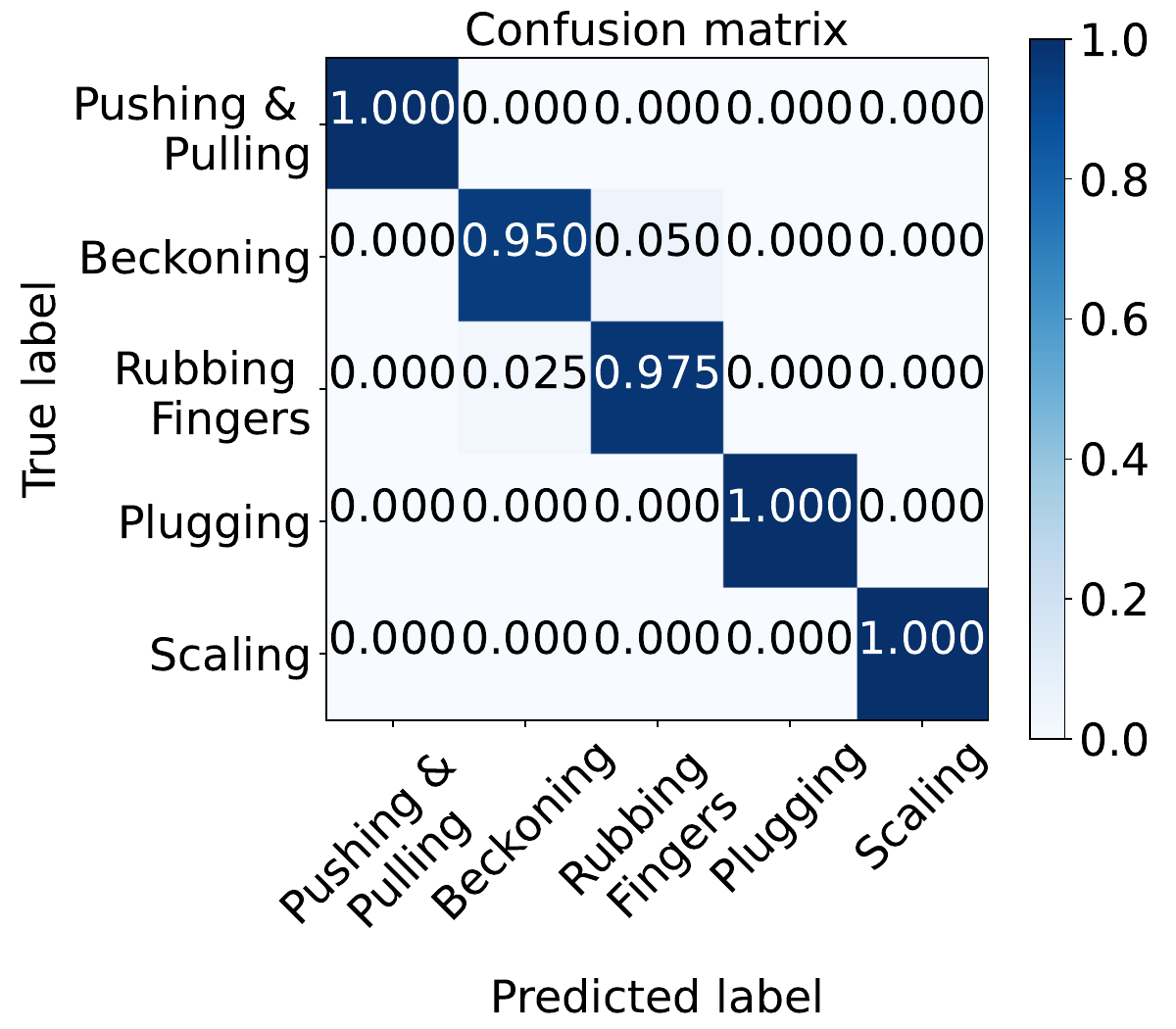}
    }
    \\
    \subfloat[Scheme $4$]{
        \includegraphics[width=\widthConfMatrix]{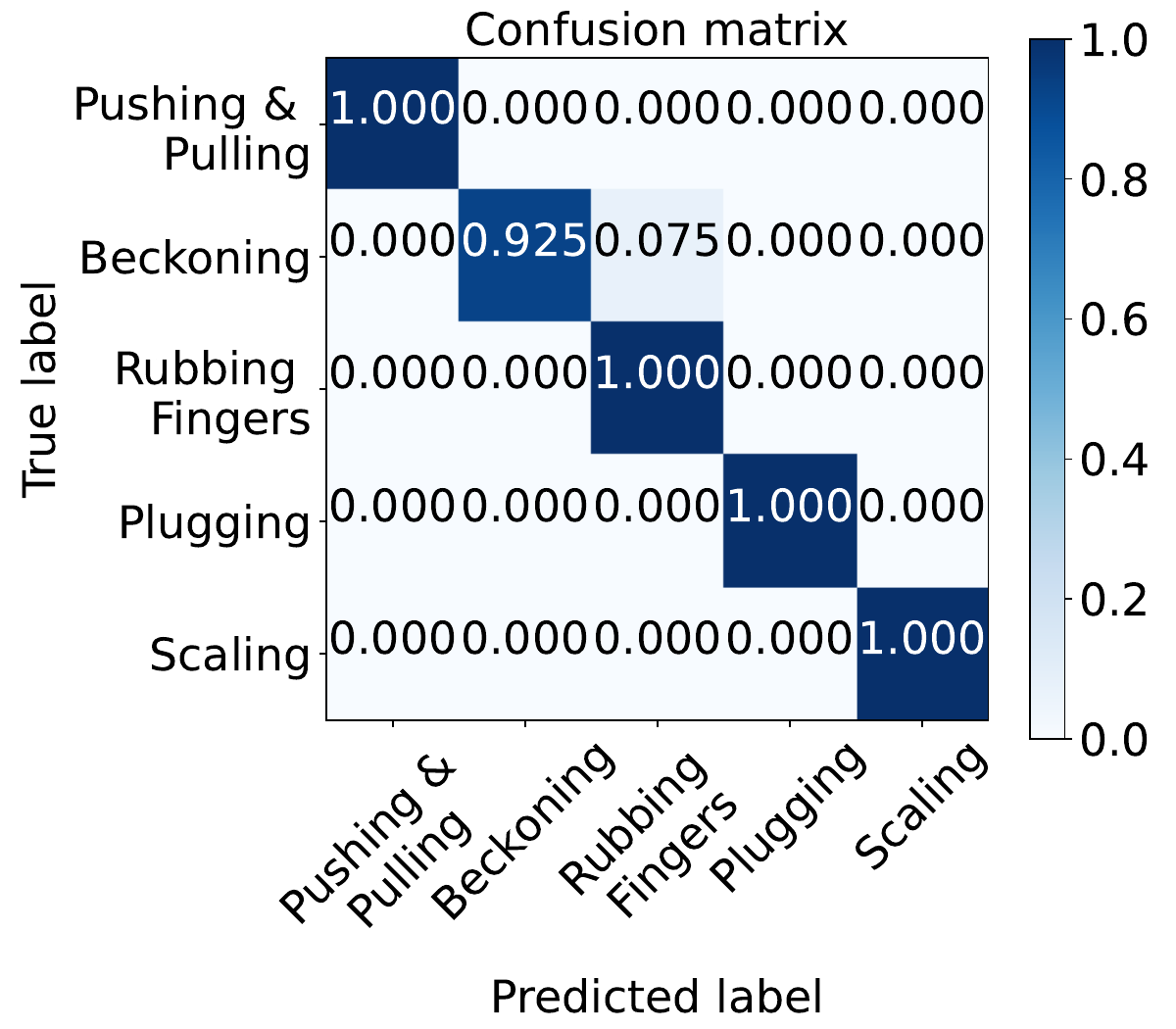}
    }
    \subfloat[Scheme $5$]{
        \includegraphics[width=\widthConfMatrix]{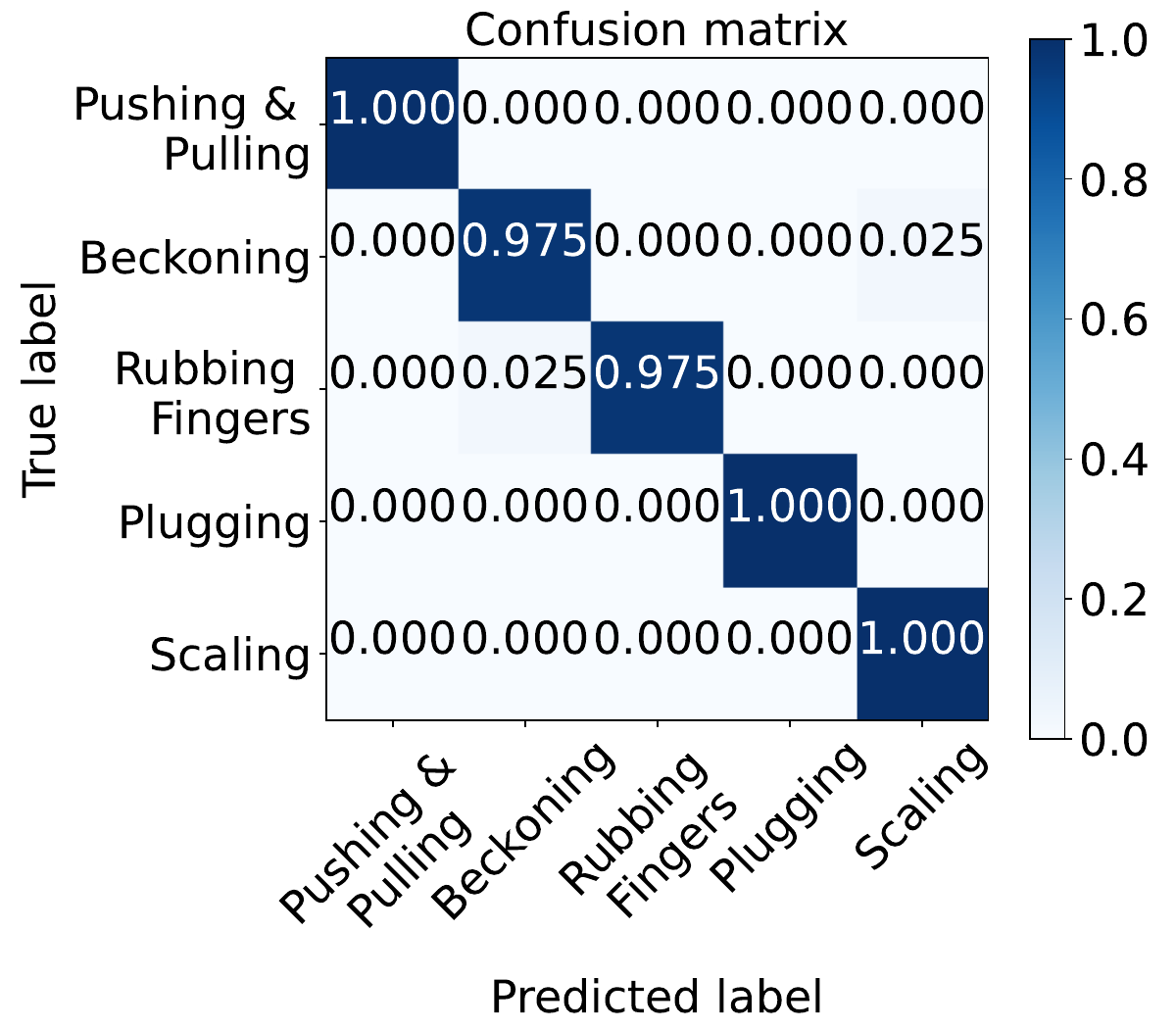}
    }
    \subfloat[Scheme $6$]{
        \includegraphics[width=\widthConfMatrix]{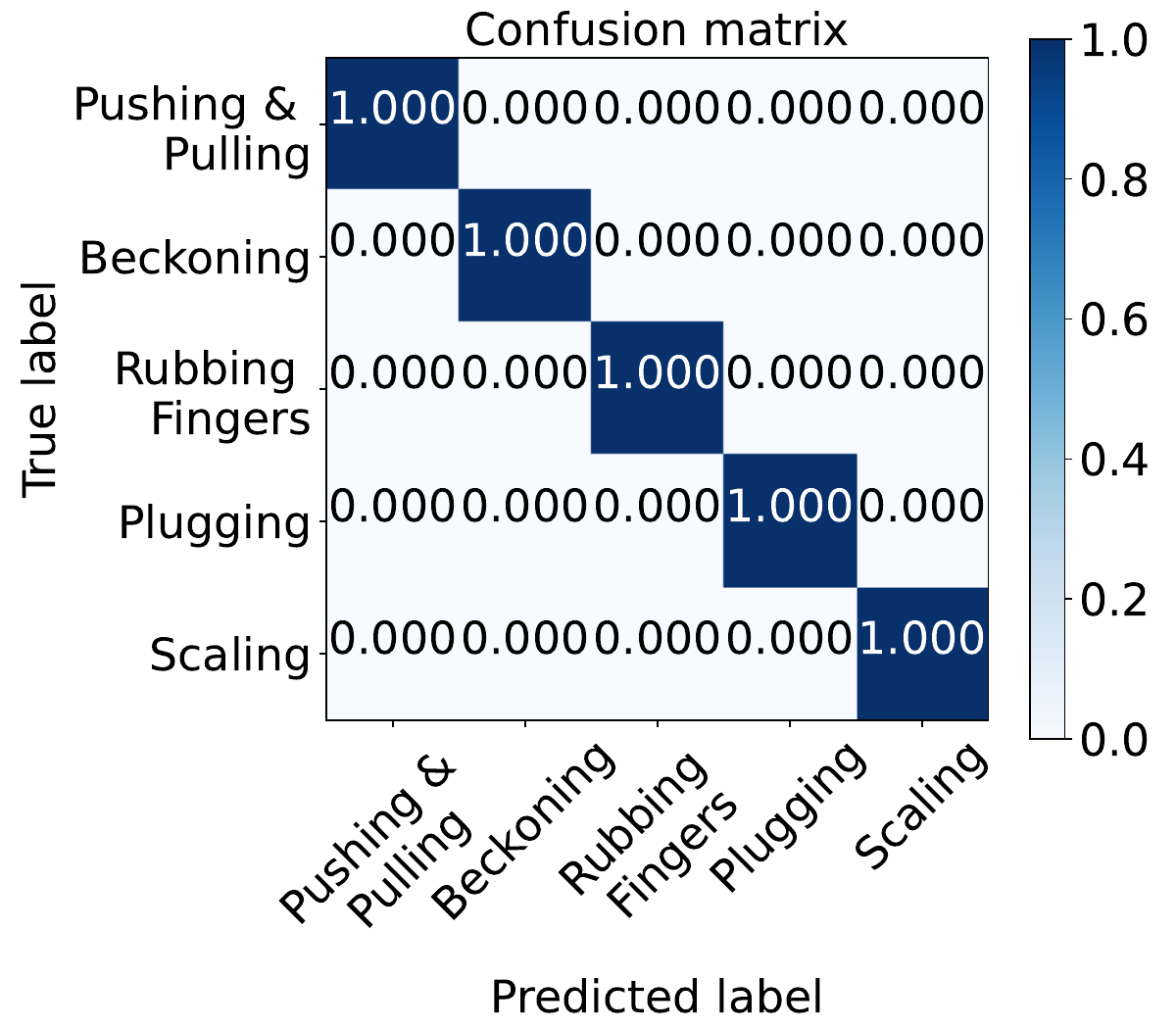}
    }
    \caption{Confusion charts of the $6$ training and testing schemes.}
    \label{fig:confusion_charts}
    % \vspace{-0.3cm}
\end{figure*}

\begin{figure}[htb]
    \centering
    \includegraphics[width=0.9\linewidth]{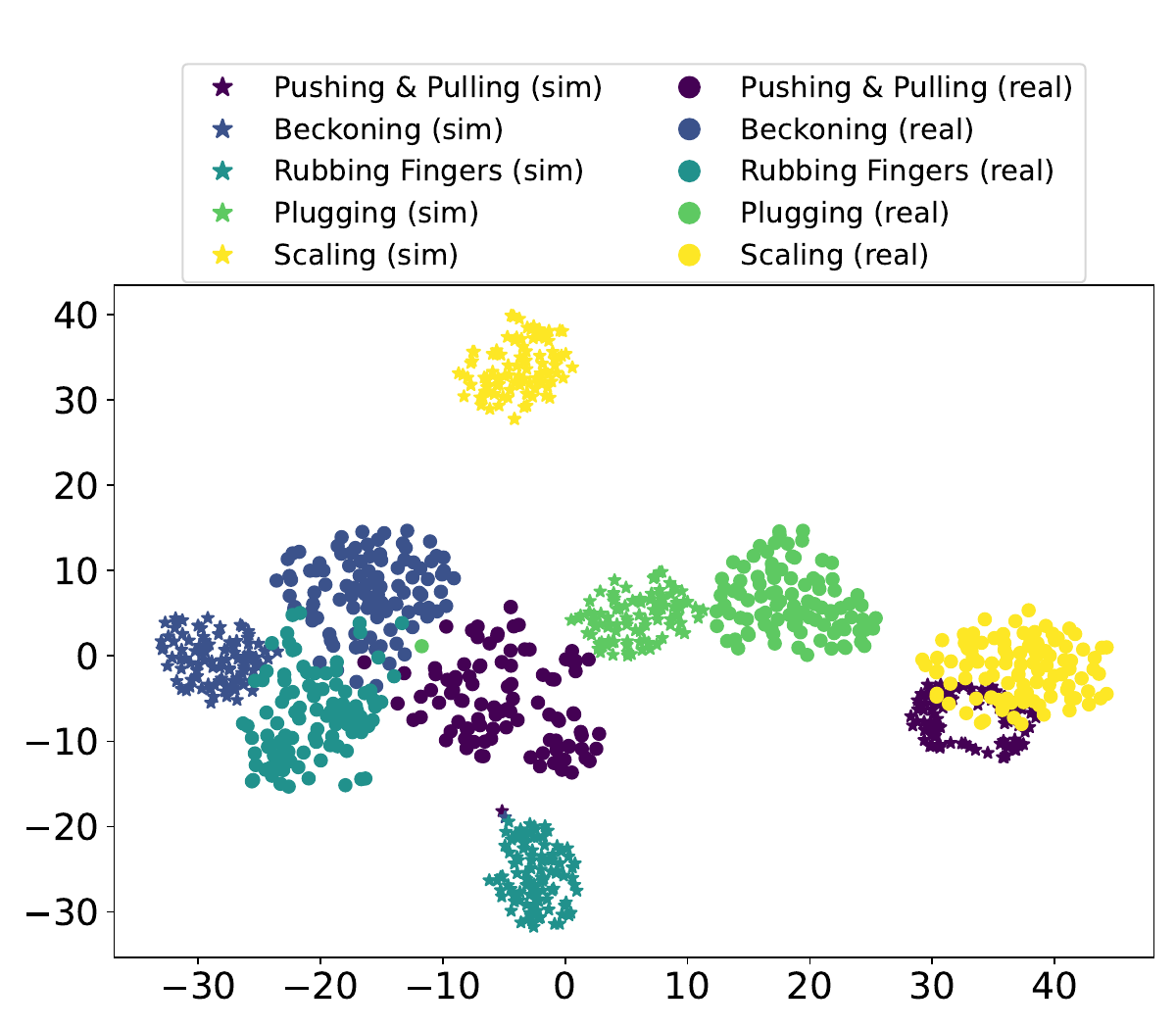}
    \caption{t-SNE visualization for the feature spaces of simulation and experimental datasets. Five gesture categories are distinguished by different colors, with simulation sample features denoted by star shapes and experimental sample features represented by solid circle shapes.}
    \label{fig:TSNE}
\end{figure}

\begin{figure*}[t]
    \centering
    \subfloat[]{
        \includegraphics[width = 0.45\linewidth]{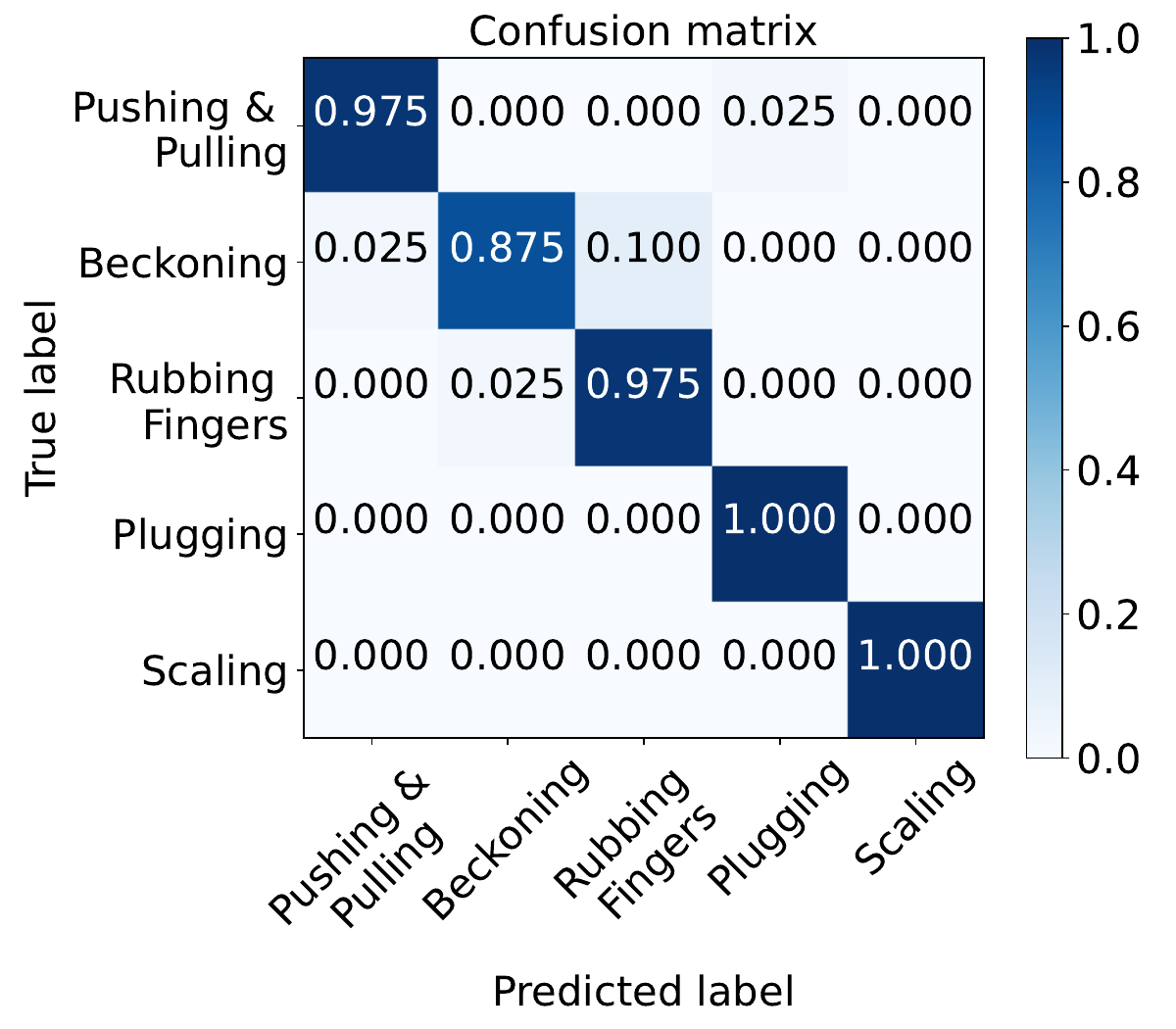}
    }
    \subfloat[]{
        \includegraphics[width=0.45\linewidth]{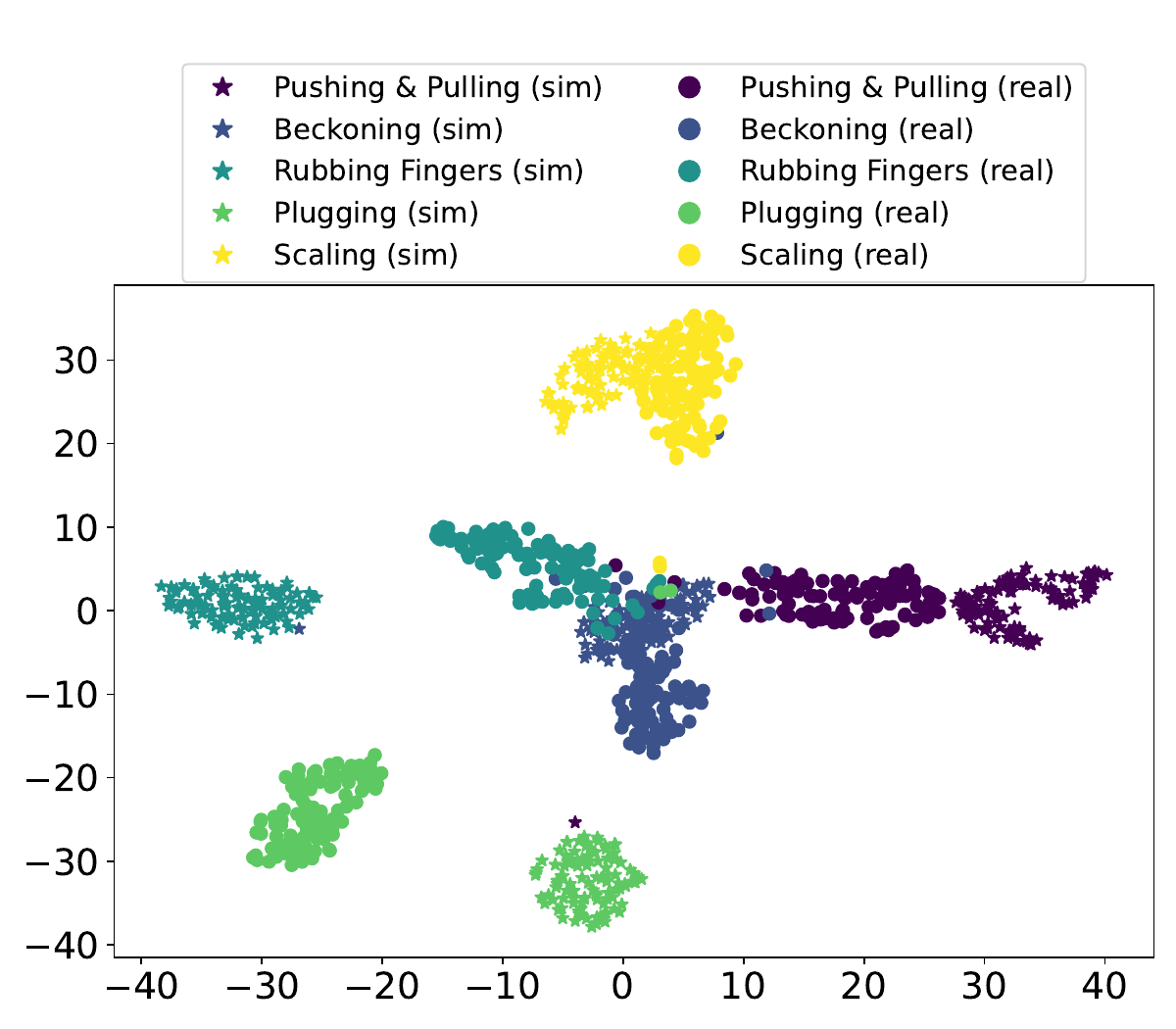}
    }
    \caption{Gesture recognition result after ADDA. (a) confusion chart of simulation-to-reality inference; (b) t-SNE visualization of the feature spaces in simulation and experimental datasets.}
    \label{fig:ADDA}
\end{figure*}

\subsection{Performance of Gesture Recognition}

First of all, it can be observed from Fig.~\ref{fig:exp-simu} that the spectrograms from real experiment and {\SystemName} simulator exhibit similar time-Doppler patterns. To further demonstrate the high fidelity of the proposed simulator in the applications of gesture recognition, the following six training and testing schemes are adopted with the same image recognition model named ResNet18\cite{he2016deep}: 
\begin{itemize}
    \item \textbf{Scheme $1$:} The training set consists of $60$ simulated spectrograms for each gesture, and the test set consists of $40$ measured ones for each gesture; 
    \item \textbf{Scheme $2$:} The training set consists of $50$ simulated spectrograms and $10$ measured ones for each gesture, and the test set consists of $40$ measured ones for each gesture; 
    \item \textbf{Scheme $3$:} The training set consists of $40$ simulated spectrograms and $20$ measured ones for each gesture, and the test set consists of $40$ measured ones for each gesture; 
    \item \textbf{Scheme $4$:} The training set consists of $30$ simulated spectrograms and $30$ measured ones for each gesture, and the test set consists of $40$ measured ones for each gesture; 
    \item \textbf{Scheme $5$:} The training set consists of $60$ measured spectrograms for each gesture, and the test set consists of $40$ measured ones for each gesture; 
    \item \textbf{Scheme $6$:} The training set consists of $60$ simulated spectrograms for each gesture, and the test set consists of $40$ simulated ones for each gesture.
\end{itemize}

The overall results of the gesture recognition are shown in Fig.~\ref{fig:training_result}, and the confusion charts of the $6$ schemes are shown in Fig.~\ref{fig:confusion_charts} respectively. It can be observed that an accuracy of $83.0\%$ (Scheme $1$) can be achieved if the simulated dataset is used for training and the experimental dataset is used for testing. On the other hand, there is still roughly $16.0\%$ and $17.0\%$ performance loss compared with the Scheme $5$ and $6$, indicating that the difference between simulated and experimental datasets is not negligible. One method to mitigate such difference is to mix some experimental samples into the simulated dataset. It can be observed from the results of Scheme $2,3,4$ that  mixing some experimental samples could significantly improve the testing accuracy. Moreover, it can be observed that the enhanced recognition accuracy converges to $98.5\%$ for Schemes $3$ and $4$. However, this is still $0.5\%$ lower than the accuracy achieved with Scheme $5$. This difference indicates the inherent feature distinctions between simulated and experimental datasets.

Furthermore, it is apparent from the Fig.~\ref{fig:confusion_charts}(a) that gesture recognition for ``beckoning'' and ``plugging'' are not sufficiently accurate. The recognition accuracy is $70\%$ and $75\%$ respectively. Moreover, $22.5\%$ confusion probability exists between the gestures of ``beckoning'' and ``pushing and pulling'', indicating that some of the simulation samples of ``beckoning'' are similar to the experimental samples of ``pushing and pulling''.

To qualitatively support the aforementioned observations, we applied dimensionality reduction techniques to the extracted features (network output before entering the fully-connected layer classifier) for the entire simulation and experimental datasets using the ResNet18 model trained by Scheme $1$. Specifically, t-distributed Stochastic Neighbor Embedding (t-SNE)\cite{van2008visualizing} and Principal Component Analysis (PCA)\cite{bishop2006pattern} were employed to visualize and analyze the high-dimensional features ($512$ dimensions for ResNet18) of the dataset, as illustrated in Fig.~\ref{fig:TSNE}. It could be observed that although $83.0\%$ gesture recognition accuracy is achieved, the distributions of the features of different gestures are not sufficiently separated. Moreover, the simulated and experimental features for the gesture ``Scaling'' are not well aligned, indicating the inherent feature distinctions between simulated and experimental datasets. These could be regarded as the limitation of the proposed simulator, since the real-world channel is more complex than the simulated one due to the impacts of multipath and the non-ideal hardware.

\subsection{Improvement via Transfer Learning}

The transfer learning technique \cite{9134370} is applied in this part to relieve the above issue of feature distinction. In this context, the simulated dataset is referred to as the source domain, and the experimental dataset as the target domain. The adversarial discriminative domain adaptation (ADDA) \cite{8099799} is adopted to align the feature distributions of the source and target domains. The ResNet18 model trained by Scheme $1$ in the previous part, serves as the source domain gesture recognition model, and the target domain gesture recognition model is initialized with the same architecture and parameters. Then, additional $50$ unlabeled experimental samples are added to the simulated dataset for Scheme $1$. This is used to train a domain discriminator to distinguish the source and target domain features and fine-tune the feature extractor part of the target model alternatively, such that the source feature representation is mimicked. The details of the ADDA method can be found in \cite{8099799}. The confusion chart of the testing result and t-SNE visualization of the  feature spaces in simulation and experimental datasets after ADDA are shown in Fig.~\ref{fig:ADDA}. The recognition accuracy is boosted to $96.5\%$ and the feature spaces of simulation and experimental datasets are well aligned. This result indicates that the feature distinctions between simulated and experimental datasets can be mitigated significantly by transfer learning.

% \vspace{-0.1cm}
\section{Conclusion}
\label{sec:conclusion}
In this paper, a computer-vision assisted wireless channel simulator, namely {\SystemName} simulator, is proposed to generate high-fidelity dataset for hand gesture recognition. In the simulator, the target hand is modeled by $21$ ellipsoid primitives, and the ray-tracing method is adopted to calculate the channel impulse responses. Moreover, a video gesture catcher is proposed to capture real motion data of gestures. In the experiments with $5$ different gestures, both real dataset via experiment and simulated dataset via {\SystemName} simulator are obtained. An accuracy of $83.0\%$ can be achieved in simulation-to-reality inference, i.e., using simulated and experimental datasets in model training and inference respectively. Moreover, this accuracy can be boosted to $96.5\%$ by transfer learning, i.e., fine-tuning the gesture recognition model with a few unlabeled real data.

% 攀登计划结题了，所以就不需要了
% \section*{Acknowledgments}
% This work was supported by the Special Funds for the Cultivation of Guangdong College Students' Scientific and Technological Innovation. ("Climbing Program" Special Funds). pdjh2023c11009.

% kpsewhich IEEEabrv.bib
% \scriptsize 
\bibliography{IEEEabrv,ref}
\bibliographystyle{IEEEtran}

\end{document}